\documentclass[12pt]{article}
\title{\bf Field Correlators in\\ 
Abelian-Projected Theories and\\
Stochastic Vacuum Model}
\author{Dmitri Antonov \thanks{Permanent address:
Institute of Theoretical and Experimental Physics, 
B. Cheremushkinskaya 25, RU-117 218 Moscow, Russia.}{\,}
\thanks{Tel.: + 39 050 844 536; Fax: + 39 050 844 538; 
E-mail address: {\tt antonov@difi.unipi.it}} 
\\
{\it INFN-Sezione di Pisa, Universit\'a degli studi di Pisa,}\\
{\it Dipartimento di Fisica, Via Buonarroti, 2 - Ed. B - 56127 Pisa, Italy}}
\date{}
\begin{document}
\maketitle
\vspace{1mm}
\centerline{\bf {Abstract}}
\vspace{3mm}
\noindent
The bilocal electric field strength correlators in  
Abelian-projected $SU(2)$- and $SU(3)$-theories
are derived with accounting for the contributions to these  
quantities brought about by the 
virtual vortex loops, built out of the dual 
Nielsen-Olesen strings. Owing to the screening 
of magnetic charge of the dual vector bosons in the gas of such loops, 
these bosons become heavier, which leads to the respective 
change of the correlation length of the vacuum in the models under study.
Besides that, it turns out that such a screening leads also to the appearance 
of the long-range contribution to one of the two coefficient functions,
which parametrize the bilocal correlator within the Stochastic Vacuum 
Model. Finally, the modifications of the propagators of the dual vector  
bosons, inspired by the correlation effects in the 
gas of vortex loops, are also discussed.

\vspace{3mm}

\newpage

\section{Introduction}

Stochastic Vacuum Model (SVM)~\cite{1} is nowadays commonly 
recognized to be
one of the most promising nonperturbative approaches to QCD
(see Ref.~\cite{2} for recent reviews).
Within the so-called bilocal or Gaussian approximation, 
well confirmed by the existing lattice data~\cite{3,4}, this model 
is fully described by the irreducible
bilocal gauge-invariant field strength correlator (cumulant),
$\left<\left<F_{\mu\nu}(x)\Phi(x,x')
F_{\lambda\rho}(x')\Phi(x',x)\right>\right>$. Here, 
$F_{\mu\nu}=\partial_\mu A_\nu-\partial_\nu A_\mu-ig[A_\mu, A_\nu]$
stands for the Yang-Mills field strength tensor, $\Phi(x,y)\equiv
\frac{1}{N_c}{\cal P}\exp\left(ig\int\limits_{y}^{x}
A_\mu(u)du_\mu\right)$ is a 
parallel transporter factor along the straight-line
path, and $\left<\left<{\cal O}{\cal O}'\right>\right>\equiv
\left<{\cal O}{\cal O}'\right>-\left<{\cal O}\right>\left<{\cal O}'\right>$
with the average defined {\it w.r.t.} the Euclidean Yang-Mills action.
It is further convenient to parametrize the bilocal cumulant by 
the two coefficient functions~\cite{1,2} as follows:

$$
\frac{g^2}{2}\left<\left<F_{\mu\nu}(x)\Phi(x,x')
F_{\lambda\rho}(x')\Phi(x',x)\right>\right>=\hat 1_{N_c\times 
N_c}\left\{(\delta_{\mu\lambda}\delta_{\nu\rho}-\delta_{\mu\rho}
\delta_{\nu\lambda})D\left((x-x')^2\right)+\right.$$

\begin{equation}
\label{dd1}
\left.+\frac12\left[\partial_\mu^x((x-x')_\lambda
\delta_{\nu\rho}-(x-x')_\rho\delta_{\nu\lambda})+
\partial_\nu^x((x-x')_\rho\delta_{\mu\lambda}-(x-x')_\lambda
\delta_{\mu\rho})\right]D_1\left((x-x')^2\right)\right\}.
\end{equation}
After that, setting for the nonperturbative parts of the $D$- and 
$D_1$-function various {\it Ans\"atze}, one can employ SVM for 
precise calculations of the processes of high-energy scattering~\cite{5}
or test these {\it Ans\"atze} in the lattice experiments~\cite{3,4}.
However from the pure field-theoretical point of view, 
it remains a great challenge to derive 
the coefficient functions analytically. Unfortunately, in this way
no big progress has up to now been achieved in the QCD itself. There
have rather been derived some relations between cumulants of various 
orders~\cite{6}, which might be useful only in testing 
the IR asymptotic behaviours of the coefficient functions. 

Contrary to that, more progress has recently been achieved 
in a derivation of the bilocal cumulant 
in some models allowing for an analytical description of confinement.
Those include Abelian-projected (AP) $SU(2)$-~\cite{th} and 
$SU(3)$~\cite{suz} theories, as well as compact QED~\cite{comp}.
The bilocal field strength cumulant in these theories has been studied 
in Refs.~\cite{7,8,9}, respectively (see Ref.~\cite{10} for a review).
The present paper also follows
this line of investigations and is devoted to the improvement of 
calculations of the bilocal cumulant in AP theories.
This improvement is based on the well known fact~\cite{11} 
that in the case of zero temperature 
under study, Abrikosov vortices~\cite{abrvor} in the Ginzburg-Landau theory 
(dual Nielsen-Olesen strings~\cite{nilol} in our 4D-case) form bound states, 
built out of a vortex and an antivortex, which are usually referred to
as vortex dipoles (vortex loops in 4D). Such vortex loops are short
living (virtual) objects, whose typical sizes are much smaller than the 
typical distances between them. This means that similarly to 
monopoles in 3D compact QED, vortex loops form a dilute gas.
The summation over the grand canonical ensemble of vortex loops 
in such a dilute gas approximation was performed in Ref.~\cite{12}
for the case of the usual Abelian Higgs model (dual 
AP $SU(2)$-gluodynamics) in 3D and 4D and then 
extended to the case of the 4D AP $SU(3)$-gluodynamics 
in Ref.~\cite{13}~\footnote{Note that in the 
case of the 2D Abelian Higgs model, such a summation has for the first 
time been performed in Ref.~\cite{14}.}. On the other hand,
in all the investigations of the bilocal field strength cumulants 
in AP theories, performed in Refs.~\cite{7,8}, 
the contribution of vortex loops to the partition function, and 
consequently to the cumulants themselves, was disregarded. 
As it will be demonstrated in the next Section, this approximation 
is really valid, since it is equivalent to another one, which states that
the typical sizes of vortex loops are negligibly small.
However, such a neglection of the contribution of vortex loops makes the 
calculations of the field strength correlators, 
performed in the above mentioned papers, essentially classical.
The improvement of these calculations, presented in this paper, is 
just based on  
accounting for the correlations in the gas of vortex loops.
Clearly, such correlations are of the quantum origin, as well as the 
grand canonical ensemble of virtual vortex loops itself.
Besides that, we shall also evaluate the contributions of vortex
loops to the propagators of the dual vector bosons and discuss the 
so-emerging modifications of the respective classical expressions.

The paper is organized as follows. In the next Section, we shall firstly 
review the main aspects of a derivation of AP theories, necessary for the 
future purposes. Secondly, we shall review the main 
results of a calculation of electric field
strength correlators in the approximation when the contribution  
of vortex loops to these quantities is disregarded. In Section 3, 
after a brief review of the properties of the grand canonical ensemble 
of vortex loops, we shall evaluate the contribution of these objects to the 
field strength correlators. In Section 4, the same will
be done for the propagators of the dual vector bosons. 
The main results of the paper are summarized in Conclusion.
In the Appendix, some technical details of the calculation
of a certain typical integral from the main text are outlined.

\section{Electric Field Strength Correlators in the Absence 
of Vortex Loops Revisited}

\subsection{The Models}

To derive from the gluodynamics 
Lagrangian the IR effective $SU(2)$- and $SU(3)$ AP
theories, based on the assumption of condensation of Cooper pairs of  
AP monopoles, one usually employs the so-called 
Abelian dominance hypothesis~\cite{15}. It states that the off-diagonal 
(in the sense of the Cartan decomposition) fields can be disregarded, since 
after the Abelian projection those can be shown to become very massive
and therefore irrelevant to the IR region, where confinement holds.
Let us start our analysis with the $SU(2)$-theory.
Then, the action describing the rest, diagonal, fields and AP
monopoles reads

\begin{equation}
\label{et}
S_{\rm eff.}\left[a_\mu, f_{\mu\nu}^{\rm m}\right]=
\frac14\int d^4x\left(f_{\mu\nu}+
f_{\mu\nu}^{\rm m}
\right)^2. 
\end{equation}
Here, $a_\mu\equiv A_\mu^3$, $f_{\mu\nu}=\partial_\mu a_\nu-\partial_\nu
a_\mu$, and 
the monopole field strength tensor $f_{\mu\nu}^{\rm m}$
obeys Bianchi identities modified by monopoles, 
$\partial_\mu\tilde f^{\rm m}_{\mu\nu}
\equiv
\frac12\varepsilon_{\mu\nu\lambda\rho}
\partial_\mu f^{\rm m}_{\lambda\rho}=j_\nu^{\rm m}$.
The monopole currents $j_\mu^{\rm m}$'s should eventually 
be averaged over in the sense, which will be specified below.

To proceed with the investigation of the monopole ensemble, 
it is reasonable to cast 
the theory under study to the dual form. This yields the following expression
for the partition function:

\begin{equation}
\label{et2}
{\cal Z}=\left<\int {\cal D}B_\mu
\exp\left[-\int d^4x\left(\frac14F_{\mu\nu}^2
-iB_\mu j_\mu^{\rm m}\right)\right]\right>_{j_\mu^{\rm m}},
\end{equation}
where $B_\mu$ is the magnetic vector-potential dual to the 
electric one, $a_\mu$, and $F_{\mu\nu}=\partial_\mu B_\nu-
\partial_\nu B_\mu$. Once the $j_\mu^{\rm m}$-dependence of the 
action became explicit, it is now possible to set up the 
properties of the monopole ensemble. To describe the condensation 
of monopole Cooper pairs, it is first necessary to specify $j_\mu^{\rm m}$
as the collective current of $N$ of those:

$$j_\mu^{{\rm m}{\,}(N)}(x)=2g_m\sum\limits_{n=1}^{N}\oint dx_\mu^n(s)
\delta(x-x^n(s)).$$
Here, the world line of the $n$-th Cooper pair is parametrized 
by the vector $x_\mu^n(s)$, and 
$g_m$ is the magnetic coupling constant, related to the QCD
coupling constant $g$ via the topological quantization condition 
$gg_m=4\pi n$ with $n$ being an integer. In what follows, 
we shall for concreteness restrict ourselves to the monopoles possessing
the minimal charge, {\it i.e.} set $n=1$, although the generalization 
to an arbitrary $n$ is straightforward.
Secondly, it is necessary to set for the measure 
$\left<\ldots\right>_{j_\mu^{\rm m}}$ the following expression~\cite{16}:

$$
\left<\exp\left(i\int d^4xB_\mu j_\mu^{\rm m}\right)\right>_{j_\mu^{\rm m}}=
1+\sum\limits_{N=1}^{\infty}\frac{1}{N!}\left[\prod\limits_{n=1}^{N}
\int\limits_{0}^{+\infty}\frac{ds_n}{s_n}{\rm e}^{4\lambda\eta^2s_n}
\int\limits_{u(0)=u(s_n)}^{}{\cal D}u(s_n')\right]\times$$

\begin{equation}
\label{et3}
\times\exp\left\{\sum\limits_{l=1}^{N}\int\limits_{0}^{s_l}ds_l'\left[
-\frac14\dot u^2(s_l')+2ig_m\dot u_\mu(s_l')B_\mu(u(s_l'))\right]-
4\lambda\sum\limits_{l,k=1}^{N}\int\limits_{0}^{s_l}ds_l'
\int\limits_{0}^{s_k}ds_k''\delta\left[u(s_l')-u(s_k'')\right]\right\}.
\end{equation}
Here, the vector $u_\mu(s_n')$ parametrizes the same contour as the 
vector $x_\mu^n(s)$. Clearly, the world-line action standing in the 
exponent on the R.H.S. of Eq.~(\ref{et3}) contains besides the usual 
free part also the term responsible for the short-range repulsion of the 
trajectories of Cooper pairs.
Equation~(\ref{et3}) can further be rewritten as an integral over the 
dual Higgs field, describing magnetic Cooper pairs, as follows:

\begin{equation}
\label{et4} 
\left<\exp\left(i\int d^4xB_\mu j_\mu^{\rm m}\right)
\right>_{j_\mu^{\rm m}}=
\int {\cal D}\Phi {\cal D}\Phi^{*}
\exp\left\{-\int d^4x\left[\frac12\left|D_\mu
\Phi\right|^2+\lambda\left(|\Phi|^2-\eta^2\right)^2\right]\right\},
\end{equation}
where $D_\mu=\partial_\mu-2ig_mB_\mu$ is the covariant 
derivative~\footnote{A seeming divergency at large proper times produced 
in Eq.~(\ref{et3}) by the factor ${\rm e}^{4\lambda\eta^2s_n}$ is 
actually apparent, since the last term in the exponent on the R.H.S. 
of this equation yields the desired damping.}. Finally, substituting 
Eq.~(\ref{et4}) into Eq.~(\ref{et2}), we arrive at the following IR 
effective AP theory of the $SU(2)$-gluodynamics:

\begin{equation}
\label{et5}
{\cal Z}=\int \left|\Phi\right| {\cal D}\left|\Phi\right|
{\cal D}\theta {\cal D}B_\mu\exp\left\{-\int d^4x\left[
\frac14 F_{\mu\nu}+
\frac12\left|D_\mu
\Phi\right|^2+\lambda\left(|\Phi|^2-\eta^2\right)^2\right]\right\},
\end{equation}
where $\Phi(x)=\left|\Phi(x)\right|{\rm e}^{i\theta(x)}$. Clearly, 
as soon as we have disregarded the off-diagonal degrees of freedom and 
demanded the condensation of monopole Cooper pairs, this theory is 
nothing else, but just the dual Abelian Higgs model.

Analogous considerations can be applied to the $SU(3)$-gluodynamics.
The only difference is that since the $SU(3)$-group has two 
diagonal generators, the resulting AP 
theory will also be $[U(1)]^2$ magnetically gauge-invariant.
Within the Abelian dominance hypothesis, the initial action 
reads

\begin{equation}
\label{pureglue}
S_{\rm eff.}\left[{\bf a}_\mu, {\bf f}_{\mu\nu}^{\rm m}\right]=
\frac14\int d^4x\left({\bf f}_{\mu\nu}+
{\bf f}_{\mu\nu}^{\rm m}\right)^2,
\end{equation}
and after the dualization we have for the partition function
({\it cf.} Eq.~(\ref{et2})):

\begin{equation}
\label{puredual}
{\cal Z}=\left<
\int {\cal D}{\bf B}_\mu
\exp\left\{-\int d^4x\left[\frac14{\bf F}_{\mu\nu}^2
-i{\bf B}_\mu {\bf j}_\mu^{\rm m}
\right]\right\}\right>_{{\bf j}_\mu^{\rm m}}.
\end{equation}
Here, ${\bf F}_{\mu\nu}=\partial_\mu {\bf B}_\nu-\partial_\nu 
{\bf B}_\mu$ 
is the field strength tensor of magnetic field ${\bf B}_\mu$, which is 
dual to the field ${\bf a}_\mu\equiv\left(A_\mu^3,A_\mu^8\right)$, 
and ${\bf j}_\nu^{\rm m}=\partial_\mu
\tilde{\bf f}_{\mu\nu}^{\rm m}$. A minor nontriviality, one meets further
{\it w.r.t.} the simplest $SU(2)$-case, is the necessity to take into 
account the fact that monopole charges are 
distributed over the lattice defined by the root vectors, which 
have the form

$$
{\bf e}_1=\left(1,0\right),~ 
{\bf e}_2=\left(-\frac12,-\frac{\sqrt{3}}{2}\right),~ 
{\bf e}_3=\left(-\frac12,\frac{\sqrt{3}}{2}\right).$$ 
These vectors 
naturally appear within the Cartan decomposition of the original 
set of gluonic fields as the 
structural constants in the commutation relations
between the diagonal and so-called step (raising and lowering) operators.
The collective current of 
$N$ monopole Cooper pairs then reads

\begin{equation}
\label{colcur}
{\bf j}_\mu^{{\rm m}{\,}(N)}(x)=2g_m
\sum\limits_{n=1}^{N}\sum\limits_{a=1}^{3}{\bf e}_a\oint 
dx_\mu^{(a){\,}n}(s)
\delta\left(x-x^{(a){\,}n}(s)\right).
\end{equation}
As far as the average over the currents is concerned, it 
has the form  

$$
\left<\exp\left(i\int d^4x{\bf B}_\mu {\bf j}_\mu^{\rm m}\right)
\right>_{{\bf j}_\mu^{\rm m}}=\prod\limits_{a=1}^{3}\left\{
1+\sum\limits_{N=1}^{\infty}\frac{1}{N!}\left[\prod\limits_{n=1}^{N}
\int\limits_{0}^{+\infty}\frac{ds_n}{s_n}{\rm e}^{4\lambda\eta^2s_n}
\int\limits_{u^{(a)}(0)=u^{(a)}(s_n)}^{}Du^{(a)}
(s_n')\right]\times\right.$$

$$
\times\exp\left[\sum\limits_{l=1}^{N}
\int\limits_{0}^{s_l}ds_l'\left(
-\frac14\left(\dot u^{(a)}(s_l')\right)^2
+2ig_m\dot u_\mu^{(a)}(s_l'){\bf e}_a
{\bf B}_\mu\left(u^{(a)}(s_l')\right)\right)-\right.
$$

$$
\left.\left.-4\lambda\sum\limits_{l,k=1}^{N}\int\limits_{0}^{s_l}ds_l'
\int\limits_{0}^{s_k}ds_k''\delta\left[u^{(a)}(s_l')-u^{(a)}(s_k'')
\right]\right]\right\}=$$

\begin{equation}
\label{glue1}
=\int {\cal D}\Phi_a {\cal D}\Phi_a^{*}\exp\left\{-\int d^4x
\sum\limits_{a=1}^{3}\left[\frac12\left|\left(\partial_\mu-
2ig_m{\bf e}_a{\bf B}_\mu\right)\Phi_a\right|^2+
\lambda\left(|\Phi_a|^2-\eta^2\right)^2\right]\right\},
\end{equation}
where the vector $u_\mu^{(a)}(s_n')$ parametrizes the 
same contour as the vector $x_\mu^{(a){\,}n}(s)$. Finally, it is 
worth noting that since monopoles are distributed over the root lattice,
whose vectors are related to each other by the condition 
$\sum\limits_{a=1}^{3}{\bf e}_a=0$, the dual Higgs fields 
$\Phi_a$'s are also not completely independent of each other.
In Ref.~\cite{suz}, it was argued that the relevant constraint
for these fields reads $\sum\limits_{a=1}^{3}\theta_a=0$. 
Taking this into account 
we arrive at the following partition function describing an effective 
$[U(1)]^2$ magnetically gauge-invariant AP theory of the 
$SU(3)$-gluodynamics~\cite{suz}:

$$
{\cal Z}=\int \left|\Phi_a\right| {\cal D}\left|\Phi_a\right|
{\cal D}\theta_a {\cal D}{\bf B}_\mu
\delta\left(\sum\limits_{a=1}^{3}
\theta_a\right)\times$$

\begin{equation}
\label{et6}
\times\exp\left\{-\int d^4x\left[\frac14{\bf F}_{\mu\nu}^2+
\sum\limits_{a=1}^{3}\left[\frac12\left|\left(\partial_\mu-
2ig_m{\bf e}_a{\bf B}_\mu\right)\Phi_a\right|^2+
\lambda\left(|\Phi_a|^2-\eta^2\right)^2\right]\right]\right\},
\end{equation}
where $\Phi_a=\left|\Phi_a\right|{\rm e}^{i\theta_a}$.
 
\subsection{Bilocal Electric Field Strength Correlators}

\subsubsection{$SU(2)$-case}

In order to investigate bilocal cumulants of electric field 
strengths in the models~(\ref{et5}) and~(\ref{et6}), it is necessary
to extend them by external electrically charged test particles
({\it i.e.} particles, charged {\it w.r.t.} the Cartan subgroup
of the original $SU(2)$- or $SU(3)$-group). For brevity, we shall call 
these particles simply ``quarks''. 
In the $SU(2)$-case,
such an extension can be performed by adding to the 
action~(\ref{et}) the term $i\int d^4xa_\mu j_\mu^{\rm e}$ with  
$j_\mu^{\rm e}(x)\equiv g\oint
\limits_{C}^{}dx_\mu(s)\delta(x-x(s))$ standing for the conserved 
electric current of a quark, which moves along a certain closed contour $C$.
Then, performing the dualization of the so-extended action
and summing up over 
monopole currents according to Eq.~(\ref{et3}), 
we arrive at Eq.~(\ref{et5}) with $F_{\mu\nu}$ replaced by
$F_{\mu\nu}+F_{\mu\nu}^{\rm e}$. Here,  
$F_{\mu\nu}^{\rm e}$ stands for the field strength tensor 
generated by quarks according to the equation $\partial_\mu\tilde 
F_{\mu\nu}^{\rm e}=j_\nu^{\rm e}$. 
A solution to this equation reads $F_{\mu\nu}^{\rm e}
=-g\tilde\Sigma_{\mu\nu}^{\rm e}$, where   
$\Sigma_{\mu\nu}^{\rm e}(x)\equiv\int\limits_{\Sigma^{\rm e}}^{}
d\sigma_{\mu\nu}(\bar x(\xi))\delta(x-\bar x(\xi))$ 
is the so-called vorticity 
tensor current defined at an arbitrary surface $\Sigma^{\rm e}$
(which is just the world sheet of an open dual 
Nielsen-Olesen string), bounded by the contour $C$.

From now on, we shall be interested 
in the London limit, $\lambda\to\infty$, of the
theories~(\ref{et5}) and~(\ref{et6}), where they admit an exact 
string representation. In that limit, the partition function 
of the theory~(\ref{et5}) with external quarks reads

\begin{equation}
\label{vosem}
{\cal Z}=\int {\cal D}B_\mu {\cal D}\theta^{{\rm sing.}} 
{\cal D}\theta^{{\rm reg.}}\exp\left\{-\int d^4x\left[\frac14
\left(F_{\mu\nu}+F_{\mu\nu}^{\rm e}\right)^2+\frac{\eta^2}{2}
\left(\partial_\mu\theta-2g_mB_\mu\right)^2\right]\right\}. 
\end{equation}
In Eq.~(\ref{vosem}), we have performed a 
decomposition of the phase of the dual 
Higgs field $\theta=
\theta^{{\rm sing.}}+\theta^{{\rm reg.}}$, where the multivalued field 
$\theta^{{\rm sing.}}(x)$ 
describes a certain configuration of the dual strings and 
obeys the equation~\cite{17}
 
\begin{equation}
\label{devyat}
\varepsilon_{\mu\nu\lambda\rho}\partial_\lambda
\partial_\rho\theta^{{\rm sing.}}(x)=2\pi\Sigma_{\mu\nu}(x). 
\end{equation}
Here, $\Sigma_{\mu\nu}$ stands for the 
vorticity tensor current, defined at the world sheet $\Sigma$ of a 
closed dual string, parametrized by the vector $x_\mu(\xi)$.
On the other hand, the field $\theta^{\rm reg.}(x)$ 
describes simply a singlevalued fluctuation around the above mentioned
string configuration. 

The string representation of the theory~(\ref{vosem}) 
can be derived similarly to Ref.~\cite{17}, where this has been done 
for a model with a global $U(1)$-symmetry. One gets 

\begin{equation}
\label{odinnad}
{\cal Z}=
\int {\cal D}x_\mu(\xi) {\cal D}h_{\mu\nu} 
\exp\Biggl\{-\int d^4x\left[\frac1{12\eta^2}H_{\mu\nu
\lambda}^2+g_m^2h_{\mu\nu}^2+i\pi h_{\mu\nu}\hat\Sigma_{\mu\nu}
\right]\Biggr\},
\end{equation}
where $\hat\Sigma_{\mu\nu}
\equiv4\Sigma_{\mu\nu}^{\rm e}-\Sigma_{\mu\nu}$, and   
$H_{\mu\nu\lambda}\equiv\partial_\mu h_{\nu\lambda}+
\partial_\lambda h_{\mu\nu}+\partial_\nu h_{\lambda\mu}$ is the field 
strength 
tensor of a massive antisymmetric tensor field $h_{\mu\nu}$ (the so-called 
Kalb-Ramond field~\cite{18}). This antisymmetric spin-1 tensor field 
emerged via some constraints from the integration over 
$\theta^{\rm reg.}$ and represents the  
massive dual vector boson. As far as the integration over the 
world sheets of closed strings, 
${\cal D}x_\mu(\xi)$, is concerned, it appeared from the 
integration over $\theta^{\rm sing.}$ by virtue of  
Eq.~(\ref{devyat}), owing to which there exists a one-to-one 
correspondence between $\theta^{\rm sing.}$ and $x_\mu(\xi)$. 
Physically this correspondence stems from the fact that the singularity 
of the phase of the dual Higgs field just takes place at the string 
world sheets. (Notice that since in what follows we shall be 
interested in effective actions rather than the integration measures, 
the Jacobian emerging during the change of the integration variables 
$\theta^{\rm sing.}\to x_\mu(\xi)$, which has been evaluated 
in Ref.~\cite{19}, will not be discussed below and is assumed 
to be included into the measure ${\cal D}x_\mu(\xi)$.)

Finally, the Gaussian 
integration over the field $h_{\mu\nu}$ in Eq.~(\ref{odinnad}) 
leads to the following expression for the  
partition function~(\ref{vosem}):

$$
{\cal Z}=\exp\left[-\frac{g^2}{2}\oint\limits_C^{}dx_\mu
\oint\limits_C^{}dy_\mu D_m^{(4)}(x-y)\right]\times$$

\begin{equation}
\label{pyatnad}
\times\int {\cal D}x_\mu(\xi)\exp\left[-(\pi\eta)^2\int d^4x\int d^4y
\hat\Sigma_{\mu\nu}(x)D_m^{(4)}(x-y)\hat\Sigma_{\mu\nu}(y)
\right]. 
\end{equation}
Here, $D_m^{(4)}(x)\equiv\frac{m}{4\pi^2|x|}K_1(m|x|)$ is the 
propagator of the dual vector boson, whose mass $m$, generated 
by the Higgs mechanism, is equal to
$2g_m\eta$, and $K_\nu$'s
henceforth stand for the modified Bessel functions. The details 
of derivation of Eqs.~(\ref{odinnad}) and~(\ref{pyatnad}) 
can be found {\it e.g.} in Ref.~\cite{10}. Besides 
that review, the obtained string 
representation~(\ref{pyatnad}) has been discussed in  
various contexts in Refs.~\cite{7,9,12,19,20}.
Clearly, the first exponential factor 
on the R.H.S. of Eq.~(\ref{pyatnad}) 
is the standard result, which can be obtained without accounting for the 
dual Nielsen-Olesen strings. Contrary to that, 
the integral over string world sheets on the R.H.S. of 
this equation stems just from the contribution of strings to the 
partition function and is 
the essence of the string representation.
The respective string effective action describes
both the interaction of the closed world sheets $\Sigma$'s with the  
open world sheets $\Sigma^{\rm e}$'s and self-interactions of these
objects.

We are now in the position to discuss the bilocal correlator 
of electric field strengths in the model~(\ref{vosem}).
Indeed, owing to the Stokes theorem,
such an extended partition function (which is actually nothing else,
but the Wilson loop of a test quark) 
can be written as $\left<\exp\left(-\frac{ig}{2}\int d^4x
\Sigma^{\rm e}_{\mu\nu}f_{\mu\nu}\right)
\right>_{a_\mu, j_\mu^{\rm m}}$, where 
$\left<\ldots\right>_{a_\mu, j_\mu^{\rm m}}\equiv\left<\int 
{\cal D}a_\mu\exp
\left(-S_{\rm eff.}\left[a_\mu, f_{\mu\nu}^{\rm m}\right]\right) 
\left(\ldots\right)\right>_{j_\mu^{\rm m}}$ with 
$S_{\rm eff.}$ and $\left<\ldots\right>_{j_\mu^{\rm m}}$
given by Eqs.~(\ref{et}) and (\ref{et3}), respectively. Applying to this 
expression the cumulant expansion, we have in the bilocal approximation:

\begin{equation}
\label{Zonehand}
{\cal Z}\simeq\exp\left[-\frac{g^2}{8}\int d^4x\int d^4y
\Sigma_{\mu\nu}^{\rm e}(x)\Sigma_{\lambda\rho}^{\rm e}(y)\left<\left<
f_{\mu\nu}(x)f_{\lambda\rho}(y)
\right>\right>_{a_\mu, j_\mu^{\rm m}}\right].
\end{equation}
Following the SVM, let us parametrize the bilocal cumulant  
by the two Lorentz structures similarly to Eq.~(\ref{dd1}): 

$$\left<\left<f_{\mu\nu}(x)f_{\lambda\rho}(0)\right>
\right>_{a_\mu, j_\mu^{\rm m}}=
\Biggl(\delta_{\mu\lambda}\delta_{\nu\rho}-\delta_{\mu\rho}
\delta_{\nu\lambda}\Biggr){\cal D}\left(x^2\right)+$$

\begin{equation}
\label{dvaddva}
+\frac12\Biggl[\partial_\mu
\Biggl(x_\lambda\delta_{\nu\rho}-x_\rho\delta_{\nu\lambda}\Biggr)
+\partial_\nu\Biggl(x_\rho\delta_{\mu\lambda}-x_\lambda\delta_{\mu\rho}
\Biggr)\Biggr]{\cal D}_1\left(x^2\right). 
\end{equation}
Owing to the Stokes theorem, Eq.~(\ref{dvaddva}) yields

\begin{equation}
\label{eventual}
{\cal Z}\simeq\exp\left\{-\frac18\int d^4x\int d^4y\left[2g^2
\Sigma_{\mu\nu}^{\rm e}(x)
\Sigma_{\mu\nu}^{\rm e}(y)
{\cal D}\left((x-y)^2\right)+
j_\mu^{\rm e}(x)j_\mu^{\rm e}(y)G\left((x-y)^2\right)
\right]\right\},
\end{equation} 
where 

\begin{equation}
\label{g}
G\left(x^2\right)\equiv\int\limits_{x^2}^{+\infty}d\lambda
{\cal D}_1(\lambda).
\end{equation}

On the other hand, Eq.~(\ref{eventual}) should coincide with 
Eq.~(\ref{pyatnad}) divided by ${\cal Z}\left[\Sigma_{\mu\nu}^{\rm e}=0
\right]$ (which is just the standard normalization 
condition, encoded in the integration measures), {\it i.e.} it reads

$${\cal Z}=\exp\left\{-\int d^4x\int d^4y D_m^{(4)}(x-y)\left[
(4\pi\eta)^2\Sigma_{\mu\nu}^{\rm e}(x)
\Sigma_{\mu\nu}^{\rm e}(y)+\frac12
j_\mu^{\rm e}(x)j_\mu^{\rm e}(y)\right]\right\}\times$$

\begin{equation}
\label{otherhand}
\times\left<\exp\left[8(\pi\eta)^2\int d^4x\int d^4y
D_m^{(4)}(x-y)\Sigma_{\mu\nu}^{\rm e}(x)
\Sigma_{\mu\nu}(y)\right]\right>_{x_\mu(\xi)},
\end{equation}
where 

\begin{equation}
\label{teqnul}
\left<\ldots\right>_{x_\mu(\xi)}\equiv\frac{\int {\cal D}x_\mu(\xi)
\left(\ldots\right)\exp\left[-(\pi\eta)^2\int d^4x\int d^4y
\Sigma_{\mu\nu}(x)D_m^{(4)}(x-y)\Sigma_{\mu\nu}(y)\right]}{
\int {\cal D}x_\mu(\xi)
\exp\left[-(\pi\eta)^2\int d^4x\int d^4y
\Sigma_{\mu\nu}(x)D_m^{(4)}(x-y)\Sigma_{\mu\nu}(y)\right]}.
\end{equation}
As it has already been discussed in the Introduction, 
in the case of zero temperature under study, dual Nielsen-Olesen
strings, one should average over in Eq.~(\ref{teqnul}), form 
virtual bound states of vortex loops. The typical areas $|\Sigma|$'s
of those are very small, and in the leading approximation 
can be disregarded {\it w.r.t.} the area $|\Sigma^{\rm e}|$ 
of the world sheet of the long open string, which confines a test quark.
Owing to this, the exponential factor, which should be averaged over 
vortex loops on the R.H.S. of Eq.~(\ref{otherhand}), can be 
disregarded {\it w.r.t.} the first exponential factor in this equation, 
as well. Then, the comparison 
of the latter one with Eq.~(\ref{eventual}) straightforwardly yields 
for the function ${\cal D}$ the following expression

\begin{equation}
\label{dvadtri}
{\cal D}\left(x^2\right)=\frac{m^3}{4\pi^2}
\frac{K_1(m|x|)}{\left|x\right|},   
\end{equation}
whereas for the function ${\cal D}_1$ we get the equation
$G\left(x^2\right)=4D_m^{(4)}(x)$, which leads to:

\begin{equation}
\label{dvadchetyr}
{\cal D}_1\left(x^2\right)=
\frac{m}{2\pi^2x^2}\Biggl[\frac{K_1(m|x|)}{\left|x\right|}
+\frac{m}{2}\Biggl(K_0(m|x|)+K_2(m|x|)\Biggr)\Biggr]. 
\end{equation}
We see that in the IR limit $\left|x\right|\gg\frac1m$, 
the asymptotic behaviours of the coefficient functions~(\ref{dvadtri}) 
and~(\ref{dvadchetyr}) 
are given by  

\begin{equation}
\label{dvadpyat}
{\cal D}\longrightarrow\frac{m^4}{4\sqrt{2}\pi^{\frac32}}
\frac{{\rm e}^{-m\left|x\right|}}{\left(m\left|x\right|\right)^
{\frac32}}  
\end{equation}
and 

\begin{equation}
\label{dvadshest}
{\cal D}_1\longrightarrow\frac{m^4}{2\sqrt{2}\pi^{\frac32}}
\frac{{\rm e}^{-m\left|x\right|}}{\left(m\left|x\right|\right)^
{\frac52}}. 
\end{equation}
For bookkeeping purposes, let us also present the asymptotic 
behaviours of these functions 
in the opposite case, $\left|x\right|
\ll\frac1m$. Those read 
 
\begin{equation}
\label{dvadsem} 
{\cal D}\longrightarrow\frac{m^2}{4\pi^2x^2}  
\end{equation}
and

\begin{equation}
\label{dvadvosem}
{\cal D}_1\longrightarrow\frac{1}{\pi^2\left|x\right|^4}. 
\end{equation}

One can now see that according to the lattice data~\cite{3,4},  
the asymptotic behaviours~(\ref{dvadpyat}) and~(\ref{dvadshest}) 
are very similar 
to the IR ones of the nonperturbative parts of the 
functions $D$ and $D_1$, which parametrize the bilocal 
cumulant~(\ref{dd1}) in the case of QCD.
In particular, both 
functions decrease exponentially, and the function ${\cal D}$ 
is much larger 
than the function ${\cal D}_1$ 
due to the preexponential power-like behaviour.
We also see that the r\^ole of the correlation length of the 
vacuum, $T_g$, is the SVM, 
{\it i.e.} the distance at which the functions $D$ and $D_1$ decrease,
is played in the model~(\ref{vosem}) by the inverse 
mass of the dual vector boson, $\frac{1}{m}$.
Moreover, the UV asymptotic behaviours~(\ref{dvadsem}) 
and~(\ref{dvadvosem}) also 
parallel the results of the SVM in QCD to 
the lowest order 
of perturbation theory. Namely, at such distances the 
function $D_1$ 
also behaves as $\frac{1}{\left|x\right|^4}$ due to the one-gluon-exchange
contribution. As far as the function $D$ is concerned, it vanishes to
the leading order of perturbation theory. Although 
this is not the case in our model (whose UV features are far from those
of the asymptotically free QCD), 
the ${\cal D}_1$-asymptotics~(\ref{dvadvosem}) is nevertheless 
really much larger than that of the 
${\cal D}$-function, given by Eq.~(\ref{dvadsem}).
  
Hence we see that within the approximation when the contribution of 
vortex loops to the partition function~(\ref{otherhand}) is disregarded
completely, the bilocal approximation to the SVM is an exact
result in the theory~(\ref{vosem}),
{\it i.e.} all the cumulants
of the orders higher than the second one vanish. Higher cumulants
naturally appear upon performing in Eq.~(\ref{otherhand})
the average~(\ref{teqnul}) over vortex loops. However, this average 
yields important modifications already on the level of the bilocal cumulant. 
Namely, as we shall see in the next Section, 
it modifies the classical expressions~(\ref{dvadtri}) 
and~(\ref{dvadchetyr}).

\subsubsection{$SU(3)$-case}

Let us now turn ourselves to the bilocal cumulant of electric 
field strength tensors in the London limit of AP
$SU(3)$-gluodynamics, where the partition function~(\ref{et6})
of this theory takes the form 

\begin{equation}
\label{suz2}
{\cal Z}=\int {\cal D}{\bf B}_\mu {\cal D}\theta_a^{\rm sing.}
{\cal D}\theta_a^{\rm reg.} \delta\left(\sum\limits_{a=1}^{3}
\theta_a\right)\exp\Biggl\{-\int d^4x\Biggl[
\frac14{\bf F}_{\mu\nu}^2+\frac{\eta^2}{2}\sum\limits_{a=1}^{3}
\left(\partial_\mu\theta_a-2g_m{\bf e}_a{\bf B}_\mu\right)^2
\Biggr]\Biggr\}. 
\end{equation}
Similarly to the $SU(2)$-case, 
in the model under study there exist   
dual Nielsen-Olesen-type strings. Due to that, 
in Eq.~(\ref{suz2}) 
we have again decomposed the total phases 
of the dual Higgs 
fields into the multivalued and singlevalued parts, $\theta_a=
\theta_a^{\rm sing.}+\theta_a^{\rm reg.}$. Here, 
the multivalued parts $\theta_a^{\rm sing.}$'s 
describe a given configuration of the dual strings of three types. 
They are related to the world sheets $\Sigma_a$'s of these strings  
via the equations 

\begin{equation}
\label{suz3}
\varepsilon_{\mu\nu\lambda\rho}\partial_\lambda\partial_\rho
\theta_a^{\rm sing.}(x)=2\pi\Sigma_{\mu\nu}^a(x)\equiv
2\pi\int\limits_{\Sigma_a}^{}d\sigma_{\mu\nu}(x_a(\xi))
\delta(x-x_a(\xi)),
\end{equation}
where $x_a\equiv x_\mu^a(\xi)$ is a four-vector parametrizing
the world sheet $\Sigma_a$. 

An external quark of a certain colour $c=R,B,G$ (red, blue, green, 
respectively) can be introduced
into the theory under study by adding to the 
initial action~(\ref{pureglue}) the interaction term $i{\bf Q}^{(c)}
\int d^4x{\bf a}_\mu j_\mu^{\rm e}$, where 
the vectors of colour charges read 

$${\bf Q}^{(R)}=\left(\frac{1}{2}, \frac{1}{2\sqrt{3}}\right),~ 
{\bf Q}^{(B)}=\left(-\frac{1}{2}, \frac{1}{2\sqrt{3}}\right),~ 
{\bf Q}^{(G)}=\left(0, -\frac{1}{\sqrt{3}}\right).
$$
These vectors are just the weights of the representation ${\bf 3}$
of ${}^{*}SU(3)$. In another words, 
those are nothing else, but the charges of quarks
{\it w.r.t.} the Cartan subgroup $[U(1)]^2$, {\it i.e.} for every $c$,
the components of ${\bf Q}^{(c)}$ are just the charges of a  
quark of the colour $c$ {\it w.r.t.} the diagonal gluons $A_\mu^3$ and 
$A_\mu^8$.

Applying further 
the Stokes theorem and the cumulant expansion in the bilocal 
approximation, we get for the partition function of the 
theory~(\ref{suz2}) 
with an external quark of the colour $c$ the following 
expression:

\begin{equation}
\label{colour}
{\cal Z}_c\simeq\exp\left[-\frac{g^2}{8}Q^{(c)i}Q^{(c)j}\int d^4x\int d^4y
\Sigma_{\mu\nu}^{\rm e}(x)\Sigma_{\lambda\rho}^{\rm e}(y)
\left<\left<f_{\mu\nu}^i(x)
f_{\lambda\rho}^j(y)\right>\right>_{{\bf a}_\mu, {\bf j}_\mu^{\rm m}}\right].
\end{equation}
Here, $i,j=1,2$ denote the $[U(1)]^2$-indices, referring to the 
Cartan generators $(T^3,T^8)$, and the average is defined as
$\left<\ldots\right>_{{\bf a}_\mu, {\bf j}_\mu^{\rm m}}
\equiv\left<\int {\cal D}{\bf a}_\mu\exp
\left(-S_{\rm eff.}\left[{\bf a}_\mu, {\bf f}_{\mu\nu}^{\rm m}
\right]\right) 
\left(\ldots\right)\right>_{{\bf j}_\mu^{\rm m}}$, where 
$S_{\rm eff.}$ and $\left<\ldots\right>_{{\bf j}_\mu^{\rm m}}$
are given by Eqs.~(\ref{pureglue}) and (\ref{glue1}), respectively.
Upon the SVM-inspired parametrization of the bilocal cumulant,  

$$\left<\left<f_{\mu\nu}^i(x)f_{\lambda\rho}^j(0)\right>
\right>_{{\bf a}_\mu, {\bf j}_\mu^{\rm m}}=\delta^{ij}\Biggl\{
\Biggl(\delta_{\mu\lambda}\delta_{\nu\rho}-\delta_{\mu\rho}
\delta_{\nu\lambda}\Biggr)\hat D\left(x^2\right)+$$

\begin{equation}
\label{colorcorrel}
+\frac12\Biggl[\partial_\mu
\Biggl(x_\lambda\delta_{\nu\rho}-x_\rho\delta_{\nu\lambda}\Biggr)
+\partial_\nu\Biggl(x_\rho\delta_{\mu\lambda}-x_\lambda\delta_{\mu\rho}
\Biggr)\Biggr]\hat D_1\left(x^2\right)\Biggr\}, 
\end{equation}
we can write for Eq.~(\ref{colour}) the following expression:

\begin{equation}
\label{Zapr}
{\cal Z}_c\simeq\exp\left\{-\frac{1}{24}\int d^4x\int d^4y\left[
2g^2\Sigma_{\mu\nu}^{\rm e}(x)\Sigma_{\mu\nu}^{\rm e}(y)
\hat D\left((x-y)^2\right)+
j_\mu^{\rm e}(x)j_\mu^{\rm e}(y)\hat G\left((x-y)^2\right)
\right]\right\}.
\end{equation}
Here, we have denoted by $\hat G$ the same function as~(\ref{g}), 
but with the replacement ${\cal D}_1\to\hat D_1$, and used the fact 
that for every $c$, $({\bf Q}^{(c)})^2=\frac{1}{3}$.

On the other hand, one can derive the string representation for 
the partition function ${\cal Z}_c$.
Indeed, the dualization of the action~(\ref{pureglue})
with the term $i{\bf Q}^{(c)}
\int d^4x{\bf a}_\mu j_\mu^{\rm e}$ added, leads to Eq.~(\ref{puredual}) 
with ${\bf F}_{\mu\nu}$ replaced by ${\bf F}_{\mu\nu}+
{\bf F}_{\mu\nu}^{(c)}$. Here, ${\bf F}_{\mu\nu}^{(c)}$ stands for the 
field strength tensor of a test quark of the colour $c$, which obeys the 
equation $\partial_\mu\tilde{\bf F}_{\mu\nu}^{(c)}=
{\bf Q}^{(c)}j_\nu^{\rm e}$
and thus can be written as ${\bf F}_{\mu\nu}^{(c)}=-g{\bf Q}^{(c)}
\tilde\Sigma_{\mu\nu}^{\rm e}$. Next, the summation over the currents
of monopole Cooper pairs 
in the sense of Eq.~(\ref{glue1}) yields Eq.~(\ref{et6}) with the 
same extension of ${\bf F}_{\mu\nu}$. In the London limit under study, 
the string representation of this theory
(see Refs.~\cite{8,21} for details) reads

$${\cal Z}_c=\int {\cal D}x_\mu^a(\xi)\delta\left(\sum\limits_{a=1}^{3}
\Sigma_{\mu\nu}^a\right)\times$$

\begin{equation}
\label{exactZ}
\times\exp\left\{-\pi^2\int d^4x\int d^4y
D_{m_B}^{(4)}(x-y)\left[\eta^2\bar\Sigma_{\mu\nu}^a(x)
\bar\Sigma_{\mu\nu}^a(y)+\frac{1}{6\pi^2}j_\mu^{\rm e}(x)
j_\mu^{\rm e}(y)\right]\right\}.
\end{equation}
Here, $m_B=\sqrt{6}g_m\eta$ is the mass of the dual vector bosons,
which they acquire due to the Higgs mechanism. We have also introduced the 
notation $\bar\Sigma_{\mu\nu}^a\equiv\Sigma_{\mu\nu}^a-
2s_a^{(c)}\Sigma_{\mu\nu}^{\rm e}$ with the following numbers $s_a^{(c)}$'s:
$s_3^{(R)}=s_2^{(B)}=
s_1^{(G)}=0$, $s_1^{(R)}=s_3^{(B)}=s_2^{(G)}=
-s_2^{(R)}=-s_1^{(B)}=-s_3^{(G)}=1$, which obey the relation
${\bf Q}^{(c)}=\frac13{\bf e}_as_a^{(c)}$. Taking into account 
that for every $c$, $\left(s_a^{(c)}\right)^2=2$, we eventually arrive 
at the following expression for the partition function ({\it cf.}
Eq.~(\ref{otherhand})):

$${\cal Z}_c=\exp\left\{-8\pi^2\int d^4x\int d^4yD_{m_B}^{(4)}(x-y)\left[
\eta^2\Sigma_{\mu\nu}^{\rm e}(x)\Sigma_{\mu\nu}^{\rm e}(y)+\frac{1}{48\pi^2}
j_\mu^{\rm e}(x)j_\mu^{\rm e}(y)\right]\right\}\times$$

\begin{equation}
\label{Zexact}
\times\left<\exp\left[(2\pi\eta)^2s_a^{(c)}\int d^4x\int d^4y
\Sigma_{\mu\nu}^a(x)D_{m_B}^{(4)}(x-y)\Sigma_{\mu\nu}^{\rm e}(y)\right]
\right>_{x_\mu^a(\xi)}
\end{equation}
with the average over vortex loops having the form

$$\left<\ldots\right>_{x_\mu^a(\xi)}\equiv\frac{
\int {\cal D}x_\mu^a(\xi)\delta\left(\sum\limits_{a=1}^{3}
\Sigma_{\mu\nu}^a\right)\left(\ldots\right)\exp\left[
-(\pi\eta)^2\int d^4x\int d^4y\Sigma_{\mu\nu}^a(x)
D_{m_B}^{(4)}(x-y)\Sigma_{\mu\nu}^a(y)\right]}{
\int {\cal D}x_\mu^a(\xi)\delta\left(\sum\limits_{a=1}^{3}
\Sigma_{\mu\nu}^a\right)\exp\left[
-(\pi\eta)^2\int d^4x\int d^4y\Sigma_{\mu\nu}^a(x)
D_{m_B}^{(4)}(x-y)\Sigma_{\mu\nu}^a(y)\right]}.$$

Comparing now Eq.~(\ref{Zapr}) with Eq.~(\ref{Zexact}), we see that 
in the approximation of very small vortex loops, 
$|\Sigma^a|\ll|\Sigma^{\rm e}|$, 
the functions $\hat D$ and 
$\hat D_1$ are given by Eqs.~(\ref{dvadtri}) 
and~(\ref{dvadchetyr}) with the replacement 
$m\to m_B$. 
Besides that, it is obvious that 
the bilocal cumulant~(\ref{colorcorrel}) is 
nonvanishing only for the gluonic field strength tensors of the same 
kind, {\it i.e.} for $i=j=1$ or $i=j=2$. 
Hence, for these diagonal cumulants,  
the vacuum of the AP $SU(3)$-gluodynamics in the London limit 
does exhibit a nontrivial correlation length  
$T_g=\frac{1}{m_B}$. 

\section{Electric Field Strength Correlators in the Gas of Vortex Loops}

\subsection{$SU(2)$-case}

To study the properties of vortex loops in 
the above considered theories, there is clearly no necessity to 
introduce external quarks.
The field strength correlators can be studied afterwards, {\it i.e.} 
already after
the summation over the grand canonical ensemble of vortex loops.
Thus, let us first consider the theory~(\ref{vosem}) with 
$F_{\mu\nu}^{\rm e}=0$. Upon the derivation of the  
string representation of such a theory, we are then left
with Eq.~(\ref{odinnad}), where $\Sigma_{\mu\nu}^{\rm e}=0$.
To study the grand canonical ensemble of vortex loops, it is 
necessary to replace $\Sigma_{\mu\nu}$ in Eq.~(\ref{odinnad}) 
by the following expression: 

\begin{equation}
\label{13new}
\Sigma_{\mu\nu}^{\rm gas}(x)=\sum\limits_{i=1}^{N}n_i\int 
d\sigma_{\mu\nu}(x_i(\xi))\delta(x-x_i(\xi)).
\end{equation}
Here, $\xi\in [0,1]\times [0,1]$ is a 2D-coordinate, and 
$n_i$'s stand for winding numbers. In what follows, we shall 
restrict ourselves to the vortex loops possessing the minimal
winding numbers, $n_i=\pm 1$. That is because, analogously to the 
3D-case~\cite{11,22}, the energy 
of a single vortex loop is known to be a quadratic function 
of its flux, owing to which the existence of two vortex loops
of a unit flux is more energetically favourable than the 
existence of one vortex loop of the double flux. 

Then, taking into account that the gas of vortex loops is dilute,
one can perform the summation over the grand canonical ensemble 
of these objects, which yields the following expression for the 
partition function~\cite{12}:

\begin{equation}
\label{14new}
{\cal Z}=\int {\cal D}h_{\mu\nu}\exp
\left\{-\int d^4x\left[\frac{1}{12\eta^2}H_{\mu\nu\lambda}^2+
g_m^2h_{\mu\nu}^2-2\zeta\cos\left(\frac{\left|h_{\mu\nu}
\right|}{\Lambda^2}\right)\right]\right\}.
\end{equation} 
Here $\left|h_{\mu\nu}\right|\equiv\sqrt{h_{\mu\nu}^2}$, and 
$\Lambda\equiv\sqrt{\frac{L}{\pi a^3}}$ is an UV momentum cutoff
with $L$ and $a$ denoting the characteristic distances between 
vortex loops and their typical sizes, respectively.
Clearly in the dilute gas approximation under study, 
$a\ll L$ and $\Lambda\gg a^{-1}$. Also in Eq.~(\ref{14new}),  
$\zeta\propto {\rm e}^{-S_0}$ stands for the fugacity
(Boltzmann factor) of a single vortex loop, which  
has the dimension $({\rm mass})^4$, with 
$S_0$ denoting the action of a single loop.

Note that the value of $S_0$ is approximately equal to $\sigma a^2$,
where we have estimated the area of the vortex loop as $a^2$, and 
$\sigma$ stands for the string tension of the loop,
{\it i.e.} its energy per unit area. This energy can be evaluated 
from the action standing in the arguments of the exponents on the R.H.S. of  
Eq.~(\ref{teqnul}) by virtue of the results of Ref.~\cite{23} and reads 

\begin{equation}
\label{logacc}
\sigma=\frac{\eta^2}{2}\int d^2t\frac{K_1(|t|)}{|t|}
\simeq\frac{\pi\eta^2}{2}\ln\left(\frac{\lambda}{g_m^2}\right).
\end{equation}
Here we have in the standard way~\cite{22} set for a characteristic 
small dimensionless quantity in the model under study the 
value $\frac{g_m}{\sqrt{\lambda}}$, which is of the order of the ratio
of $m$ to the mass of the dual Higgs field. Moreover, it has 
been assumed that not only $\frac{\sqrt{\lambda}}{g_m}\gg 1$, but also
$\ln\left(\frac{\sqrt{\lambda}}{g_m}\right)\gg 1$, {\it i.e.} 
the last equality on the R.H.S. of Eq.~(\ref{logacc}) is valid with 
the logarithmic accuracy. The physical origin of this logarithmic 
divergency is analogous to that, 
which takes place in 3D~\cite{11,22} and is based on the fact
that at the world sheet of a vortex loop the condensate 
of the dual Higgs field is destroyed, and the dual vector boson 
remains massless.

The square of the 
full mass of the field $h_{\mu\nu}$ following from Eq.~(\ref{14new}) 
reads $M^2=m^2+m_D^2\equiv Q^2\eta^2$. Here,    
$m_D=\frac{2\eta\sqrt{\zeta}}{\Lambda^2}$ is the additional 
contribution, emerging due to the screening of magnetic charge of 
the dual vector boson in the gas of electric vortex loops, and 
$Q=2\sqrt{g_m^2+\frac{\zeta}{\Lambda^4}}$ 
is the full magnetic charge of the dual vector boson. 

To study the correlation functions of vortex loops, it is convenient 
to represent the partition function~(\ref{14new}) directly as 
an integral over these objects. This can be done 
by virtue of the following equality,

$$
\exp\left\{-\int d^4x\left[\frac{1}{12\eta^2}H_{\mu\nu\lambda}^2+
g_m^2h_{\mu\nu}^2\right]\right\}=$$

\begin{equation}
\label{sovsemnew}
=\int {\cal D}S_{\mu\nu}\exp\left\{-\left[(\pi\eta)^2
\int d^4x\int d^4y S_{\mu\nu}(x)D_m^{(4)}(x-y)
S_{\mu\nu}(y)+i\pi\int d^4x h_{\mu\nu}S_{\mu\nu}\right]\right\},
\end{equation}
in whose derivation it has been taken into account 
that $\partial_\mu h_{\mu\nu}=0$.
Indeed, owing to the Hodge decomposition theorem, the Kalb-Ramond
field can always be represented as follows: $h_{\mu\nu}=\partial_\mu
\varphi_\nu-\partial_\nu\varphi_\mu+\varepsilon_{\mu\nu\lambda\rho}
\partial_\lambda\psi_\rho$. Clearly, in the original expression for the 
partition function,

\begin{equation}
\label{inigas}
{\cal Z}=
\left<\int {\cal D}h_{\mu\nu} 
\exp\Biggl\{-\int d^4x\left[\frac1{12\eta^2}H_{\mu\nu
\lambda}^2+g_m^2h_{\mu\nu}^2-i\pi h_{\mu\nu}\Sigma_{\mu\nu}^{\rm gas}
\right]\Biggr\}\right>_{\rm gas},
\end{equation}
the field $\varphi_\mu$ decouples 
not only from $\Sigma_{\mu\nu}^{\rm gas}$ (due to the conservation 
of the latter one), but also from $\psi_\mu$. The $\varphi_\mu$-field
thus yields only an inessential determinant factor, which 
is not of our interest. Therefore this field can be disregarded, 
which proves the above statement.

Substituting now Eq.~(\ref{sovsemnew}) 
into Eq.~(\ref{14new}), we can integrate the field 
$h_{\mu\nu}$ out. This yields the desired representation 
for the partition function:

\begin{equation}
\label{15new}
{\cal Z}=\int {\cal D}S_{\mu\nu}\exp\left\{
-\left[(\pi\eta)^2
\int d^4x\int d^4y S_{\mu\nu}(x)D_m^{(4)}(x-y)
S_{\mu\nu}(y)+V[S_{\mu\nu}]\right]\right\},
\end{equation}
where the effective potential of vortex loops, $V$, reads

\begin{equation}
\label{potloops}
V[S_{\mu\nu}]=\int d^4x\left\{\pi\Lambda^2|S_{\mu\nu}|\ln\left[
\frac{\pi\Lambda^2}{2\zeta}|S_{\mu\nu}|+\sqrt{1+\left(
\frac{\pi\Lambda^2}{2\zeta}|S_{\mu\nu}|\right)^2}\right]-2\zeta
\sqrt{1+\left(
\frac{\pi\Lambda^2}{2\zeta}|S_{\mu\nu}|\right)^2}\right\}.
\end{equation}
It is straightforward to prove that the 
correlation functions of $S_{\mu\nu}$'s, 
calculated by virtue of the representation~(\ref{15new}), are nothing 
else, but the correlation functions of vortex loops in the gas. 
This can 
be seen in the following way. Let us integrate the $h_{\mu\nu}$-field 
out of the 
initial expression~(\ref{inigas}) for the partition function 
of the gas of vortex loops. This yields ({\it cf.} Eq.~(\ref{teqnul})):

\begin{equation}
\label{intermedi}
{\cal Z}=\left<\exp\left[-(\pi\eta)^2\int d^4x\int d^4y
\Sigma_{\mu\nu}^{\rm gas}(x)D_m^{(4)}(x-y)\Sigma_{\mu\nu}^{\rm gas}(y)
\right]\right>_{\rm gas}.
\end{equation}
This equation is now perfect to involve $S_{\mu\nu}$'s and demonstrate
that their correlation functions are indeed equal to those of 
vortex loops. After that, the $h_{\mu\nu}$-dependence can be restored back,
so that we shall eventually again arrive at Eq.~(\ref{14new}) with
the substitution~(\ref{sovsemnew}). In order to involve $S_{\mu\nu}$'s, 
let us rewrite Eq.~(\ref{intermedi}) as follows:

$$
{\cal Z}=1+\sum\limits_{N=1}^{\infty}\frac{\zeta^N}{N!}\times
$$

\begin{equation}
\label{Smunus}
\times
\left<\int {\cal D}S_{\mu\nu}\delta\left(S_{\mu\nu}-
\Sigma_{\mu\nu}^{\rm gas}\right)
\exp\left[-(\pi\eta)^2\int d^4x\int d^4y
S_{\mu\nu}(x)D_m^{(4)}(x-y)S_{\mu\nu}(y)
\right]\right>_{\{x_i(\xi)\}_{i=1}^{N}}.
\end{equation}
The average here reads 

\begin{equation}
\label{avia}
\left<{\cal O}\right>_{\{x_i(\xi)\}_{i=1}^{N}}
\equiv\prod\limits_{i=1}^{N}
\int d^4y_i\int {\cal D}z_i(\xi)\mu\left[z_i\right]
\sum\limits_{n_i=\pm 1}^{}{\cal O}.
\end{equation}
In this formula, the vector $y_i$ 
describes the position of the 
world sheet of the $i$-th vortex loop~\footnote{For brevity, we 
omit the Lorentz index.}, whereas the vector $z_i(\xi)$ 
describes its shape, {\it i.e.}
$x_i(\xi)=y_i+z_i(\xi)$, $y_i=\int d^2\xi
x_i(\xi)$. We have also denoted by $\mu$ a certain 
rotation- and translation invariant measure of integration 
over shapes of the world sheets of vortex loops. Note that it was just 
this average, which in the dilute gas approximation 
led from Eq.~(\ref{inigas}) to Eq.~(\ref{14new})
(see Ref.~\cite{12} for details).

From the $\delta$-function in Eq.~(\ref{Smunus}) it is now clearly seen 
that the correlation functions of 
$S_{\mu\nu}$'s are indeed equal to those of vortex loops. 
One can further represent this $\delta$-function 
as an integral over the Lagrange multiplier, whose r\^ole, as we shall
see immediately below, is just
played by the Kalb-Ramond field:

\begin{equation}
\label{Smuplus}
\delta\left(S_{\mu\nu}-
\Sigma_{\mu\nu}^{\rm gas}\right)=\int {\cal D}h_{\mu\nu}
\exp\left[-i\pi\int d^4x h_{\mu\nu}\left(S_{\mu\nu}-
\Sigma_{\mu\nu}^{\rm gas}\right)\right].
\end{equation}
Indeed, normalizing the measure ${\cal D}h_{\mu\nu}$ 
by the condition~\footnote{Owing to Eq.~(\ref{sovsemnew}), 
this condition can be rewritten simply as 
$\int {\cal D}h_{\mu\nu}
\exp\left\{-\int d^4x\left[\frac{1}{12\eta^2}H_{\mu\nu\lambda}^2+
g_m^2h_{\mu\nu}^2\right]\right\}=1$.}

$$
\int {\cal D}h_{\mu\nu}{\cal D}S_{\mu\nu}
\exp\left[-(\pi\eta)^2\int d^4x\int d^4y
S_{\mu\nu}(x)D_m^{(4)}(x-y)S_{\mu\nu}(y)-i\pi\int d^4x
h_{\mu\nu}S_{\mu\nu}\right]=1,$$
we get

$$
{\cal Z}=\int {\cal D}S_{\mu\nu}{\cal D}h_{\mu\nu}\exp 
\exp\left\{-\left[(\pi\eta)^2
\int d^4x\int d^4y S_{\mu\nu}(x)D_m^{(4)}(x-y)
S_{\mu\nu}(y)+\right.\right.$$

$$\left.\left.+i\pi\int d^4x h_{\mu\nu}S_{\mu\nu}
-2\zeta\cos\left(\frac{\left|h_{\mu\nu}
\right|}{\Lambda^2}\right)
\right]\right\}.$$
This is just Eq.~(\ref{14new}) with the substitution~(\ref{sovsemnew}),
which completes our proof.

The correlation functions of vortex loops 
can now be calculated in the approximation when the loop gas  
is sufficiently dilute, namely its density obeys the inequality
$\left|S_{\mu\nu}\right|\ll\frac{\zeta}{\Lambda^2}$. 
Within this approximation, the potential~(\ref{potloops}) becomes 
a simple quadratic functional of $S_{\mu\nu}$'s, 
and the 
generating functional for the correlators of vortex loops
takes a simple Gaussian form. It reads

$${\cal Z}[J_{\mu\nu}]=\frac{1}{{\cal Z}[0]}
\int {\cal D}S_{\mu\nu} 
\exp\left\{
-\left[(\pi\eta)^2
\int d^4x\int d^4y S_{\mu\nu}(x)D_m^{(4)}(x-y)
S_{\mu\nu}(y)+\right.\right.$$

\begin{equation}
\label{ss4d}
\left.\left.+\int d^4x\left(-2\zeta+
\frac{\pi^2\Lambda^4}{4\zeta}S_{\mu\nu}^2
+J_{\mu\nu}S_{\mu\nu}\right)\right]\right\}
=\exp\left[-\int d^4x\int d^4yJ_{\mu\nu}(x){\cal G}(x-y)J_{\mu\nu}(y)
\right],
\end{equation}
where 

\begin{equation}
\label{calGM}
{\cal G}(x)\equiv\frac{\zeta}{\pi^2\Lambda^4}(\partial^2-m^2)
D_M^{(4)}(x).
\end{equation}
Next, since $\partial_\mu\Sigma_{\mu\nu}^{\rm gas}=0$, the 
$\delta$-function in Eq.~(\ref{Smunus}) 
requires that $\partial_\mu S_{\mu\nu}=0$ 
as well. The Hodge decomposition theorem then leads to the 
following representation for $S_{\mu\nu}$: $S_{\mu\nu}=
\varepsilon_{\mu\nu\lambda\rho}\partial_\lambda\varphi_\rho$.
Owing to this fact and the same theorem, the coupling $\int d^4x
J_{\mu\nu}S_{\mu\nu}$ will be nonvanishing only provided that 
$J_{\mu\nu}=\varepsilon_{\mu\nu\lambda\rho}\partial_\lambda
I_\rho$. This coupling then reads $2\int d^4x I_\mu T_{\mu\nu}
\varphi_\nu$, where $T_{\mu\nu}(x)\equiv\partial_\mu^x\partial_\nu^x-
\delta_{\mu\nu}\partial^{x{\,}2}$.
On the other hand, substituting the above representation 
for $J_{\mu\nu}$ into the R.H.S. of Eq.~(\ref{ss4d}), we have 

$${\cal Z}[J_{\mu\nu}]=\exp\left[-2\int d^4x\int d^4y I_\mu(x)
I_\nu(y)T_{\mu\nu}(x){\cal G}(x-y)\right].$$
Thus, varying ${\cal Z}[J_{\mu\nu}]$ twice {\it w.r.t.} $I_\mu$
and setting then $I_\mu=0$, we get

$$T_{\mu\nu}(x)T_{\lambda\rho}(y)\left<\varphi_\nu(x)
\varphi_\rho(y)\right>=-T_{\mu\lambda}(x){\cal G}(x-y).$$
Due to the rotation- and translation invariance 
of space-time, it is natural to seek for 
$\left<\varphi_\nu(x)\varphi_\rho(y)\right>$
in the form of the following {\it Ansatz}: $\delta_{\nu\rho}
g(x-y)$. This yields the equation $\partial^2g={\cal G}$, whose 
solution reads 

$$g(x)=-\frac{\zeta}{\pi^2\Lambda^4}(\partial^{x{\,}2}-m^2)
\int d^4yD_0^{(4)}(x-y)D_M^{(4)}(y),$$
where $\left.D_0^{(4)}(x)\equiv D_m^{(4)}(x)\right|_{m=0}=
\frac{1}{4\pi^2x^2}$. The last integral can obviously be 
rewritten as 

\begin{equation}
\label{xzinte}
\int d^4z D_0^{(4)}(z)D_M^{(4)}(z-x).
\end{equation}
As we shall see
below, it will be necessary to know the more general expression,
namely that for the integral 

\begin{equation}
\label{mMint}
\int d^4z D_m^{(4)}(z)D_M^{(4)}(z-x).
\end{equation}
Its calculation is outlined in the Appendix, and the result 
reads

\begin{equation}
\label{Mmres}
\frac{1}{m_D^2}\left(D_m^{(4)}(x)-D_M^{(4)}(x)\right).
\end{equation}
Note that according to Eq.~(\ref{mMint}), Eq.~(\ref{Mmres}) 
should be invariant {\it w.r.t.} the interchange $m\leftrightarrow M$.
By noting that during this interchange $m_D^2$ changes its sign, 
one can see that this invariance really holds.
 
Setting now in Eq.~(\ref{Mmres}) $m=0$, we get~\footnote{
Clearly,  
this result can also be obtained directly by making use of the method 
presented in the Appendix, which was done in Ref.~\cite{21}.}  
$\frac{1}{m_D^2}
\left(D_0^{(4)}(x)-D_{m_D}^{(4)}(x)\right)$,
which yields for the 
desired integral~(\ref{xzinte}) the same result with the 
substitution $m_D\to M$. 
Thus, the final expression for the function $g$ reads

\begin{equation}
\label{newg}
g(x)=\frac{\zeta}{(\pi M\Lambda^2)^2}(\partial^2-m^2)\left(
D_M^{(4)}(x)-D_0^{(4)}(x)\right).
\end{equation}
The desired correlator of $S_{\mu\nu}$'s 
has the form 

$$\left<S_{\mu\nu}(x)S_{\lambda\rho}(y)\right>=
\varepsilon_{\mu\nu\alpha\beta}\varepsilon_{\lambda\rho\gamma\sigma}
\partial_\alpha^x\partial_\gamma^y\left<\varphi_\beta(x)\varphi_\sigma(y)
\right>$$
and therefore  

$$
\left<S_{\mu\nu}(x)S_{\lambda\rho}(0)\right>=
-\varepsilon_{\mu\nu\alpha\beta}\varepsilon_{\lambda\rho\gamma\beta}
\partial_\alpha^x\partial_\gamma^x g(x)=$$

\begin{equation}
\label{SScor}
=\left(\delta_{\lambda\nu}\delta_{\mu\rho}-\delta_{\nu\rho}
\delta_{\mu\lambda}\right){\cal G}(x)+\left(\delta_{\mu\lambda}
\partial_\rho\partial_\nu+\delta_{\nu\rho}\partial_\mu\partial_\lambda-
\delta_{\mu\rho}\partial_\lambda\partial_\nu-\delta_{\lambda\nu}
\partial_\mu\partial_\rho\right)g(x),
\end{equation}
where it has been used that $\partial^2g(x)={\cal G}(x)$.

This result can now straightforwardly be used for the calculation of the
contribution of vortex loops to the bilocal cumulant~(\ref{dvaddva}).
Indeed, applying to the average on the R.H.S. of Eq.~(\ref{otherhand}) 
the cumulant expansion in the bilocal approximation, we get: 

$${\cal Z}\simeq\exp\Biggl\{-\int d^4x\int d^4y D_m^{(4)}(x-y)\left[
(4\pi\eta)^2\Sigma_{\mu\nu}^{\rm e}(x)\Sigma_{\mu\nu}^{\rm e}(y)+\frac12
j_\mu^{\rm e}(x)j_\mu^{\rm e}(y)\right]+$$

$$+32(\pi\eta)^4\int d^4xd^4yd^4zd^4uD_m^{(4)}(x-z)D_m^{(4)}(y-u)
\Sigma_{\mu\nu}^{\rm e}(x)\Sigma_{\lambda\rho}^{\rm e}(y)
\left<\left<\Sigma_{\mu\nu}(z)
\Sigma_{\lambda\rho}(u)\right>\right>_{x_\mu(\xi)}\Biggr\}.$$
Comparing this expression with Eq.~(\ref{Zonehand}), we see that 
owing to Eq.~(\ref{SScor}), 
the additional 
contribution of vortex loops to the cumulant~(\ref{dvaddva}) has the form 

$$\Delta\left<\left<f_{\mu\nu}(x)f_{\lambda\rho}(y)\right>
\right>_{a_\mu, j_\mu^{\rm m}}=\left(4\pi g_m\eta^2\right)^2
\int d^4z\int d^4u
D_m^{(4)}(x-z)D_m^{(4)}(y-u)\times$$

$$\times\left\{\left(\delta_{\mu\lambda}\delta_{\nu\rho}-
\delta_{\mu\rho}\delta_{\nu\lambda}\right){\cal G}(z-u)+
\left[\delta_{\mu\rho}\partial_\lambda^z\partial_\nu^z+
\delta_{\nu\lambda}\partial_\mu^z\partial_\rho^z-\delta_{\mu\lambda}
\partial_\rho^z\partial_\nu^z-\delta_{\nu\rho}\partial_\mu^z
\partial_\lambda^z\right]g(z-u)\right\}.$$
Let us further compare this intermediate result 
with Eq.~(\ref{dvaddva}) and take into account that 

$$(x-y)_\mu{\cal D}_1\left((x-y)^2\right)=
-\frac12\partial_\mu^x G\left((x-y)^2\right),$$ 
where the function $G$ 
is defined by Eq.~(\ref{g}). This leads to
the following system of equations, which determine the 
contributions of vortex loops to the functions ${\cal D}$ and $G$:

\begin{equation}
\label{deltaD}
\Delta{\cal D}\left((x-y)^2\right)=\left(4\pi g_m\eta^2\right)^2
\int d^4z\int d^4uD_m^{(4)}(x-z)D_m^{(4)}(y-u){\cal G}(z-u),
\end{equation}

\begin{equation}
\label{deltaG}
\Delta G\left((x-y)^2\right)=\left(8\pi g_m\eta^2\right)^2
\int d^4z\int d^4uD_m^{(4)}(x-z)D_m^{(4)}(y-u)g(z-u).
\end{equation}
Inserting now Eq.~(\ref{calGM}) into Eq.~(\ref{deltaD}), we get

$$\Delta{\cal D}\left((x-y)^2\right)=-\frac{\left(4g_m\eta^2\right)^2
\zeta}{\Lambda^4}\int d^4uD^{(4)}_m(y-u)D^{(4)}_M(x-u).$$ 
By virtue of the Appendix, this yields

$$\Delta{\cal D}\left(x^2\right)=\frac{m^2}{4\pi^2}\left[
\frac{M}{|x|}K_1(M|x|)-\frac{m}{|x|}K_1(m|x|)\right].$$
Adding this result to Eq.~(\ref{dvadtri}), 
we finally obtain for the finction ${\cal D}$ the following full result:

\begin{equation}
\label{Dtot}
{\cal D}^{\rm full}\left(x^2\right)=\frac{m^2M}{4\pi^2}
\frac{K_1(M|x|)}{|x|}.
\end{equation}
Analogously, inserting Eq.~(\ref{newg}) into Eq.~(\ref{deltaG}),
we have 

$$\Delta G\left((x-y)^2\right)=\zeta\left(\frac{8g_m\eta^2}{\Lambda^2M}
\right)^2\int d^4uD_m^{(4)}(y-u)\left[D_0^{(4)}(x-u)-D_M^{(4)}(x-u)
\right],$$ 
or further by virtue of the Appendix, 

$$\Delta G\left(x^2\right)=\left(\frac{m_D}{\pi M|x|}\right)^2+
\left(\frac{2m}{M}\right)^2D_M^{(4)}(x)-4D_m^{(4)}(x).$$ 
Together with Eq.~(\ref{dvadchetyr}), 
this yields the following full result for 
the function ${\cal D}_1$:

\begin{equation}
\label{D1tot}
{\cal D}_1^{\rm full}\left(x^2\right)=\frac{m_D^2}{\pi^2M^2|x|^4}+
\frac{m^2}{2\pi^2Mx^2}\left[\frac{K_1(M|x|)}{|x|}+\frac{M}{2}
\left(K_0(M|x|)+K_2(M|x|)\right)\right].
\end{equation}

It is worth noting that the functions $\Delta{\cal D}$ and 
$\Delta{\cal D}_1$ contain the terms exactly equal to 
Eqs.~(\ref{dvadtri}) and (\ref{dvadchetyr}), respectively, but with the 
opposite sign,
which just cancel out in the full 
functions~(\ref{Dtot}) and (\ref{D1tot}).
We also see that, as it should be, the functions~(\ref{Dtot}) and 
(\ref{D1tot}) go over into Eqs.~(\ref{dvadtri}) and (\ref{dvadchetyr}),
respectively, when $m_D\to 0$, {\it i.e.} when one neglects the effect of 
screening in the ensemble of vortex loops. An obvious important 
consequence of the obtained Eqs.~(\ref{Dtot}) and (\ref{D1tot}) is that
the correlation length of the vacuum, $T_g$, becomes modified 
from $\frac{1}{m}$ (according to Eqs.~(\ref{dvadtri}) and (\ref{dvadchetyr}))
to $\frac{1}{M}$. (It is worth emphasizing once more that this effect is just 
due to the Debye screening of magnetic charge of the dual vector boson 
in the ensemble of electrically charged vortex loops, which makes 
this particle more heavy, namely enlarges its mass from $m$ to $M$.)
Indeed, it is straightforward to see that at $|x|\gg\frac{1}{M}$,

$${\cal D}^{\rm full}\longrightarrow\frac{(mM)^2}{4\sqrt{2}
\pi^{\frac32}}\frac{{\rm e}^{-M|x|}}{(M|x|)^{\frac32}}$$
and

$${\cal D}_1^{\rm full}\longrightarrow
\frac{m_D^2}{\pi^2M^2|x|^4}+\frac{(mM)^2}{2\sqrt{2}\pi^{\frac32}}
\frac{{\rm e}^{-M|x|}}{(M|x|)^{\frac52}}.$$
It is also remarkable  
that the leading term of the IR asymptotics 
of the function ${\cal D}_1^{\rm full}$ is a pure power-like one, rather 
than that of the function ${\cal D}_1$, given by Eq.~(\ref{dvadshest}).
Another nontrivial result is that the screening does not change the 
UV asymptotic behaviours of the functions~(\ref{dvadtri}) 
and (\ref{dvadchetyr}), {\it i.e.} the UV asymptotics of 
the functions~(\ref{Dtot}) and (\ref{D1tot}) are given by 
Eqs.~(\ref{dvadsem}) and (\ref{dvadvosem}), respectively.

Finally, it is worth remarking that due to the modification of the 
${\cal D}$-function, one could expect the appearance of some change 
in the string tension of the open dual string world sheet $\Sigma^{\rm e}$. 
However, by virtue of the general formula
expressing the string tension via the ${\cal D}$-function~\cite{23},
$\sigma=4T_g^2\int d^2z{\cal D}\left(z^2\right)$, one can check 
that this is not the case, {\it i.e.} the string tension of $\Sigma^{\rm e}$
is independent of whether we account for screening in 
the gas of vortex loops or not. The reason for that becomes clear from 
the resulting expression for $\sigma$. It reads
$16\pi\eta^2\ln\frac{1}{c}$ with $c$ standing for a characteristic small
dimensionless quantity, and thus depends 
only on $\eta$, which is not affected by screening. Similarly to 
Eq.~(\ref{logacc}), setting for $c$ the value 
$\frac{g_m}{\sqrt{\lambda}}$, we see that the string tension 
of $\Sigma^{\rm e}$ is in the factor 16 larger 
than the string tension of a vortex loop. Clearly, that is due to the 
factor 4 standing in the linear combination of 
$\Sigma_{\mu\nu}$ and $\Sigma_{\mu\nu}^{\rm e}$
in $\hat\Sigma_{\mu\nu}$ ({\it cf.} Eq.~(\ref{pyatnad})). However, the 
coupling constant of the next-to-leading term in the gradient 
expansion of the nonlocal string effective action standing in the 
second exponential factor on the R.H.S. of Eq.~(\ref{pyatnad})
(the so-called rigidity term) does depend explicitly on the 
magnetic coupling constant and therefore changes due to the screening.
Indeed, by virtue of the results of Refs.~\cite{23,10}, one can see that
for the same world sheet $\Sigma^{\rm e}$, this coupling constant 
without taking screening into account reads $\frac{2\pi}{(2g_m)^2}$,
whereas in the presence of screening it goes over to $\frac{2\pi}{Q^2}$,
as it could be intuitively expected.

\subsection{$SU(3)$-case}

In the present Subsection, we shall extend the above results 
concerning the electric 
field strength correlators in the gas of vortex loops 
to the case of AP $SU(3)$-gluodynamics~(\ref{suz2}).
When the dual Nielsen-Olesen  
strings in this theory are considered as noninteracting objects, 
the string representation of its partition function
is given by Eq.~(\ref{exactZ}), where one should set 
$\Sigma_{\mu\nu}^{\rm e}=0$. Integrating out the 
coordinates of one of the three 
world sheets (for concreteness, $x_\mu^3(\xi)$), we can write the 
so-obtained expression for the partition function as

$$
{\cal Z}=\int {\cal D}x_\mu^1(\xi) {\cal D}x_\mu^2(\xi)\times
$$

\begin{equation}
\label{5su3}
\times\exp\left\{
-2(\pi\eta)^2\int d^4x\int d^4y\left[\Sigma_{\mu\nu}^1(x)
\Sigma_{\mu\nu}^1(y)+\Sigma_{\mu\nu}^1(x)\Sigma_{\mu\nu}^2(y)+
\Sigma_{\mu\nu}^2(x)\Sigma_{\mu\nu}^2(y)\right]
D_{m_B}^{(4)}(x-y)\right\}.
\end{equation}
In order to proceed from the individual strings to the grand canonical 
ensemble of vortex loops, one should replace 
$\Sigma_{\mu\nu}^a(x)$, $a=1,2$, in Eq.~(\ref{5su3}) 
by 

$$
\Sigma_{\mu\nu}^{a{\,}{\rm gas}}(x)=\sum\limits_{i=1}^{N}n_i^a
\int d\sigma_{\mu\nu}\left(x_i^a(\xi)\right)\delta\left(
x-x_i^a(\xi)\right).
$$
Here, $n_i^a$'s stand for winding numbers, which 
we shall again set to be equal $\pm 1$. 
Performing such a replacement, one can see the crucial difference 
of the grand canonical ensemble of vortex loops in the model 
under study from that in the AP
$SU(2)$-gluodynamics, studied in the previous Subsection. 
Namely, the system 
has now the form of two interacting gases consisting of the vortex loops 
of two kinds, whereas in the $SU(2)$-case the gas was built out of 
vortex loops of the only one kind.

Analogously to that case, we shall 
treat such a grand canonical ensemble of vortex loops in the dilute 
gas approximation. According to it, characteristic sizes of loops are 
much smaller than characteristic distances between them, which in 
particular means that the vortex loops are short living (virtual) objects.
Then the summation over this grand canonical ensemble
can be most easily performed by inserting the 
unity 

\begin{equation}
\label{auxsu3}
1=\int {\cal D}S_{\mu\nu}^a\delta\left(S_{\mu\nu}^a-
\Sigma_{\mu\nu}^{a{\,}{\rm gas}}\right)
\end{equation} 
into the R.H.S. of Eq.~(\ref{5su3}) (where 
$\Sigma_{\mu\nu}^a$ is replaced by $\Sigma_{\mu\nu}^{a{\,}{\rm gas}}$)  
and representing the $\delta$-function as an integral over the Lagrange 
multiplier ({\it cf.} Eqs.~(\ref{Smunus}) and (\ref{Smuplus})).
Then, the contribution of $N$ vortex loops of each kind 
to the grand canonical ensemble takes the following form:  

$$
{\cal Z}\left[\Sigma_{\mu\nu}^{a{\,}{\rm gas}}\right]
=\int {\cal D}S_{\mu\nu}^a {\cal D}\lambda_{\mu\nu}^a\times$$

$$
\times\exp\left\{-2(\pi\eta)^2
\int d^4x\int d^4y\left[S_{\mu\nu}^1(x)
S_{\mu\nu}^1(y)+S_{\mu\nu}^1(x)S_{\mu\nu}^2(y)+
S_{\mu\nu}^2(x)S_{\mu\nu}^2(y)\right]
D_{m_B}^{(4)}(x-y)-\right.$$

\begin{equation}
\label{6su3}
\left.-i\int d^4x\lambda_{\mu\nu}^a\left(S_{\mu\nu}^a-
\Sigma_{\mu\nu}^{a{\,}{\rm gas}}\right)\right\}.
\end{equation}
After that, the desired summation over the ensemble of loops 
is straightforward, since it technically 
parallels the one of AP $SU(2)$-gluodynamics. 
We have 

$$
\prod\limits_{a=1}^{2}\left[
1+\sum\limits_{N=1}^{\infty}\frac{\zeta^N}{N!}\left<
\exp\left(i\int d^4x\lambda_{\mu\nu}^a
\Sigma_{\mu\nu}^{a{\,}{\rm gas}}\right)\right>_{
\{x_i^a(\xi)\}_{i=1}^{N}}\right]=
$$

\begin{equation}
\label{7su3}
=\exp\left\{2\zeta\int d^4y\left[\cos\left(\frac{\left|
\lambda_{\mu\nu}^1(y)\right|}{\Lambda^2}\right)+
\cos\left(\frac{\left|
\lambda_{\mu\nu}^2(y)\right|}{\Lambda^2}\right)\right]\right\},
\end{equation}
where for every $a$, the average $\left<\ldots\right>_{
\{x_i^a(\xi)\}_{i=1}^{N}}$ is given by Eq.~(\ref{avia}).
Here, it has been naturally assumed that the vortex loops of 
different kinds have the same fugacity $\zeta\propto {\rm e}^{-S_0}$,
since different $\theta_a^{\rm sing.}$'s enter the initial 
partition function~(\ref{suz2}) in the symmetric way.
Clearly, the action $S_0$ of a single loop 
can be estimated analogously to how it has been done in the 
previous Subsection for the $SU(2)$-case.
In Eq.~(\ref{7su3}), 
we have also introduced a new UV momentum cutoff $\Lambda\equiv\sqrt{
\frac{L}{a^3}}$ $\left(\gg a^{-1}\right)$, 
where $a$ again denotes a typical size of the 
vortex loop, whereas $L$ stands for a typical distance between loops, so that 
in the dilute gas approximation under study $a\ll L$. Finally in 
Eq.~(\ref{7su3}), we have denoted $\left|\lambda_{\mu\nu}^a\right|\equiv
\sqrt{\left(\lambda_{\mu\nu}^a\right)^2}$. 
 
Next, it is possible to integrate out the Lagrange multipliers 
by solving the saddle-point equation following from Eqs.~(\ref{6su3}) 
and~(\ref{7su3}), 

$$\frac{\lambda_{\mu\nu}^a}{\left|\lambda_{\mu\nu}^a\right|}\sin\left(
\frac{\left|\lambda_{\mu\nu}^a\right|}{\Lambda^2}\right)=
-\frac{i\Lambda^2}{2\zeta}S_{\mu\nu}^a.$$
After that, we 
arrive at the following representation for the partition function 
of the grand canonical ensemble of vortex loops:

$$
{\cal Z}=
\int {\cal D}S_{\mu\nu}^a
\exp\left\{-\left[2(\pi\eta)^2
\int d^4x\int d^4y\left[S_{\mu\nu}^1(x)
S_{\mu\nu}^1(y)+S_{\mu\nu}^1(x)S_{\mu\nu}^2(y)+
S_{\mu\nu}^2(x)S_{\mu\nu}^2(y)\right]\times
\right.\right.$$

\begin{equation}
\label{8su3}
\left.\left.\times D_{m_B}^{(4)}(x-y)
+V\left[\frac{S_{\mu\nu}^1}{\pi}\right]+
V\left[\frac{S_{\mu\nu}^2}{\pi}\right]\right]\right\},
\end{equation}
which owing to Eq.~(\ref{auxsu3}) is natural to be referred to as 
the representation in terms of the vortex loops. In Eq.~(\ref{8su3}), 
the effective potential of vortex loops is given by 
Eq.~(\ref{potloops}).

Next, to get the Debye masses, corresponding to the two 
interacting gases of vortex loops, it is necessary to get 
the respective sine-Gordon theory of the two Kalb-Ramond fields.
In order to derive it, let us first 
diagonalize the quadratic form in square brackets 
on the R.H.S. of Eq.~(\ref{6su3}), which can be done upon the 
introduction of the new integration variables 
${\cal S}_{\mu\nu}^1=\frac{\sqrt{3}}{2}\left(S_{\mu\nu}^1+S_{\mu\nu}^2
\right)$ and ${\cal S}_{\mu\nu}^2=\frac12\left(S_{\mu\nu}^1-S_{\mu\nu}^2
\right)$. After that, Eqs.~(\ref{6su3}) 
and~(\ref{7su3}) yield

$${\cal Z}=\int {\cal D}{\cal S}_{\mu\nu}^a
{\cal D}\lambda_{\mu\nu}^a
\exp\left\{-2(\pi\eta)^2
\int d^4x\int d^4y
{\cal S}_{\mu\nu}^a(x)D_{m_B}^{(4)}(x-y)
{\cal S}_{\mu\nu}^a(y)+
\right.$$

\begin{equation}
\label{10su3}
\left.+2\zeta\int d^4x\left[\cos\left(\frac{\left|
\lambda_{\mu\nu}^1(x)\right|}{\Lambda^2}\right)+
\cos\left(\frac{\left|
\lambda_{\mu\nu}^2(x)\right|}{\Lambda^2}\right)\right]
-i\int d^4xh_{\mu\nu}^a{\cal S}_{\mu\nu}^a\right\}.
\end{equation}
Here, we have introduced the two Kalb-Ramond fields as the 
following linear combinations of the Lagrange multipliers: 
$h_{\mu\nu}^1=\frac{1}{\sqrt{3}}
\left(\lambda_{\mu\nu}^1+\lambda_{\mu\nu}^2\right)$ and 
$h_{\mu\nu}^2=\lambda_{\mu\nu}^1-\lambda_{\mu\nu}^2$. The partition 
function of the desired sine-Gordon theory can then be obtained 
from Eq.~(\ref{10su3}) by making use of the following 
equality:

$$
\int  {\cal D}{\cal S}_{\mu\nu}^a
\exp\left\{-\left[2(\pi\eta)^2
\int d^4x\int d^4y
{\cal S}_{\mu\nu}^a(x)D_{m_B}^{(4)}(x-y)
{\cal S}_{\mu\nu}^a(y)+
i\int d^4x h_{\mu\nu}^a{\cal S}_{\mu\nu}^a\right]\right\}=
$$

\begin{equation}
\label{calSh}
=\exp\left\{-\frac{1}{2\pi^2}\int d^4x\left[\frac{1}{12\eta^2}
\left(H_{\mu\nu\lambda}^a\right)^2+\frac32 
g_m^2\left(h_{\mu\nu}^a\right)^2
\right]\right\}.
\end{equation}
Similarly to the analogous equality~(\ref{sovsemnew}), 
the equality~(\ref{calSh}) can easily be proved
by noting that due to the Hodge decomposition theorem and
the equation $\partial_\mu{\cal S}_{\mu\nu}^a=0$ (which follows 
from Eq.~(\ref{auxsu3}) and conservation of 
$\Sigma_{\mu\nu}^{a{\,}{\rm gas}}$), $\partial_\mu h_{\mu\nu}^a=0$. 
Substituting further Eq.~(\ref{calSh}) into Eq.~(\ref{10su3})
and performing the rescaling $\frac{h_{\mu\nu}^a}{\pi\sqrt{2}}
\to h_{\mu\nu}^a$, 
we arrive at the following representation for the partition function 
of the grand canonical ensemble of vortex loops in terms of the 
local sine-Gordon theory, equivalent to the 
nonlocal theory~(\ref{8su3}): 

$${\cal Z}=\int {\cal D}h_{\mu\nu}^a\exp\left\{
-\int d^4x\left[
\frac{1}{12\eta^2}
\left(H_{\mu\nu\lambda}^a\right)^2+\frac32 
g_m^2\left(h_{\mu\nu}^a\right)^2
-\right.\right.$$

\begin{equation}
\label{11su3}
\left.\left.-2\zeta\left[\cos\left(\frac{\pi}{\Lambda^2\sqrt{2}}
\left|\sqrt{3}
h_{\mu\nu}^1+h_{\mu\nu}^2\right|\right)+ 
\cos\left(\frac{\pi}{\Lambda^2\sqrt{2}}\left|\sqrt{3}
h_{\mu\nu}^1-h_{\mu\nu}^2\right|\right)\right]\right]\right\}.
\end{equation}
The full masses of 
the Kalb-Ramond fields
can now be read off from Eq.~(\ref{11su3}) by 
expanding the cosines up to the quadratic terms. The result reads 
$M_a^2=m_B^2+m_a^2\equiv Q_a^2\eta^2$, 
where $m_1=\frac{2\pi\eta}{\Lambda^2}
\sqrt{3\zeta}$, $m_2=\frac{2\pi\eta}{\Lambda^2}\sqrt{\zeta}$
are the Debye masses, and we have introduced the full magnetic charges 
$Q_1=\sqrt{6g_m^2+\frac{12\pi^2\zeta}{\Lambda^4}}$, 
$Q_2=\sqrt{6g_m^2+\frac{4\pi^2\zeta}{\Lambda^4}}$.  

Next, Eq.~(\ref{8su3}) can be used for the evaluation 
of correlators of vortex loops, which due to Eq.~(\ref{auxsu3})  
are nothing else but the correlators of $S_{\mu\nu}^a$'s. 
Those are again calculable in the approximation of a dilute gas of 
vortex loops, $\left|S_{\mu\nu}^a
\right|\ll\frac{\zeta}{\Lambda^2}$.
Within this approximation, the generating functional
for correlators of $S_{\mu\nu}^a$'s reads

$${\cal Z}\left[J_{\mu\nu}^a\right]=
\int {\cal D}{\cal S}_{\mu\nu}^a
\exp\left\{
-\left[2(\pi\eta)^2
\int d^4x\int d^4y
{\cal S}_{\mu\nu}^a(x)D_{m_B}^{(4)}(x-y)
{\cal S}_{\mu\nu}^a(y)+
\right.\right.
$$

$$
\left.\left.
+\int d^4x\left[-4\zeta+
\frac{\Lambda^4}{2\zeta}\left(\frac13\left({\cal S}_{\mu\nu}^1
\right)^2+\left({\cal S}_{\mu\nu}^2\right)^2\right)+
{\cal S}_{\mu\nu}^1\frac{J_{\mu\nu}^{+}}{\sqrt{3}}
+{\cal S}_{\mu\nu}^2J_{\mu\nu}^{-}\right]
\right]\right\},$$
where $J_{\mu\nu}^a$ is a source of $S_{\mu\nu}^a$, and 
$J_{\mu\nu}^{\pm}\equiv J_{\mu\nu}^1\pm J_{\mu\nu}^2$. 
Apart from an inessential constant factor (which can as usual
be referred to the integration measure and eventually drops out
during the calculation of the correlation functions), we thus get 
for the generating functional the following expression:

$${\cal Z}\left[J_{\mu\nu}^a\right]
=\exp\left\{-\int d^4x\int d^4y
\left[J_{\mu\nu}^{+}(x)J_{\mu\nu}^{+}(y){\cal G}_1(x-y)+
J_{\mu\nu}^{-}(x)J_{\mu\nu}^{-}(y){\cal G}_2(x-y)\right]\right\},$$
where ${\cal G}_a(x)\equiv\frac{\zeta}{2\Lambda^4}
\left(\partial^2-m_B^2\right)D_{M_a}^{(4)}(x)$.
Owing to the conservation of $S_{\mu\nu}^a$'s and the 
Hodge decomposition theorem, we again have the following 
representation for $S_{\mu\nu}^a$: $S_{\mu\nu}^a=
\varepsilon_{\mu\nu\lambda\rho}\partial_\lambda\varphi_\rho^a$.
Therefore, due to the same theorem, 
$J_{\mu\nu}^a=\varepsilon_{\mu\nu\lambda\rho}\partial_\lambda I_\rho^a$, 
which yields

$${\cal Z}\left[J_{\mu\nu}^a\right]=\exp\Biggl\{-2\int d^4x\int d^4y
\Biggl[I_\mu^a(x)I_\nu^a(y)T_{\mu\nu}(x)\left({\cal G}_1(x-y)+
{\cal G}_2(x-y)\right)+$$

$$+2I_\mu^1(x)I_\nu^2(y)T_{\mu\nu}(x)
\left({\cal G}_1(x-y)-{\cal G}_2(x-y)\right)\Biggr]\Biggr\}.$$
On the other hand, the coupling $\int d^4xJ_{\mu\nu}^aS_{\mu\nu}^a$
can be written as $2\int d^4xI_\mu^aT_{\mu\nu}\varphi_\nu^a$. Thus, 
varying ${\cal Z}\left[J_{\mu\nu}^a\right]$ twice {\it w.r.t.} 
$I_\mu^a$'s and setting then $I_\mu^a=0$, 
we arrive at the following system of equations: 

$$
T_{\mu\nu}(x)T_{\lambda\rho}(y)\left<\varphi_\nu^1(x)
\varphi_\rho^1(y)\right>=
T_{\mu\nu}(x)T_{\lambda\rho}(y)\left<\varphi_\nu^2(x)
\varphi_\rho^2(y)\right>=
-T_{\mu\lambda}(x)\left({\cal G}_1(x-y)+{\cal G}_2(x-y)\right),$$

$$
T_{\mu\nu}(x)T_{\lambda\rho}(y)\left<\varphi_\nu^1(x)
\varphi_\rho^2(y)\right>=
-T_{\mu\lambda}(x)\left({\cal G}_1(x-y)-{\cal G}_2(x-y)\right).$$
Adopting for the correlators of $\varphi_\mu^a$'s the following 
{\it Ans\"atze},

$$
\left<\varphi_\nu^1(x)
\varphi_\rho^1(0)\right>=
\left<\varphi_\nu^2(x)
\varphi_\rho^2(0)\right>=\delta_{\nu\rho}f_{+}(x),~~
\left<\varphi_\nu^1(x)
\varphi_\rho^2(0)\right>=\delta_{\nu\rho}f_{-}(x),$$
we get:

$$f_{\pm}(x)=\frac{\zeta}{2\Lambda^4}\left(\partial^2-m_B^2\right)
\left[\frac{1}{M_1^2}\left(D_{M_1}^{(4)}(x)-D_0^{(4)}(x)\right)\pm
\frac{1}{M_2^2}\left(D_{M_2}^{(4)}(x)-D_0^{(4)}(x)\right)\right].$$
This result makes the choice of notations ``$f_{\pm}(x)$'' quite 
natural. Finally, the desired correlators of vortex loops read

$$\left<S_{\mu\nu}^1(x)S_{\lambda\rho}^1(0)\right>=
\left<S_{\mu\nu}^2(x)S_{\lambda\rho}^2(0)\right>=
\left(\delta_{\nu\lambda}\delta_{\mu\rho}-\delta_{\nu\rho}
\delta_{\mu\lambda}\right)\left({\cal G}_1(x)+{\cal G}_2(x)\right)+$$

\begin{equation}
\label{ssaa}
+\left(\delta_{\mu\lambda}\partial_\rho\partial_\nu+
\delta_{\nu\rho}\partial_\mu\partial_\lambda-
\delta_{\mu\rho}\partial_\lambda\partial_\nu-
\delta_{\nu\lambda}\partial_\mu\partial_\rho\right)f_{+}(x),
\end{equation}

$$
\left<S_{\mu\nu}^1(x)S_{\lambda\rho}^2(0)\right>=
\left(\delta_{\nu\lambda}\delta_{\mu\rho}-\delta_{\nu\rho}
\delta_{\mu\lambda}\right)\left({\cal G}_1(x)-{\cal G}_2(x)\right)+$$

\begin{equation}
\label{ssab}
+\left(\delta_{\mu\lambda}\partial_\rho\partial_\nu+
\delta_{\nu\rho}\partial_\mu\partial_\lambda-
\delta_{\mu\rho}\partial_\lambda\partial_\nu-
\delta_{\nu\lambda}\partial_\mu\partial_\rho\right)f_{-}(x).
\end{equation}

This result can now be applied to the calculation of
the contribution to the correlator~(\ref{colorcorrel}), brought about  
by the vortex loops. 
Indeed, applying to the average over vortex loops, standing 
on the R.H.S. of Eq.~(\ref{Zexact}) the cumulant expansion in the 
bilocal approximation, we have due to Eq.~(\ref{colour}): 

$$\Delta\left<\left<f_{\mu\nu}^i(x)f_{\lambda\rho}^i(y)\right>
\right>_{{\bf a}_\mu, {\bf j}_\mu^M}=$$

$$=-24\pi^2g_m^2\eta^4s_a^{(c)}
s_b^{(c)}\int d^4z\int d^4u D_{m_B}^{(4)}(x-z)D_{m_B}^{(4)}(y-u)
\left<S_{\mu\nu}^a(z)S_{\lambda\rho}^b(u)\right>.$$
Taking further into account the equalities

$$\left<S_{\mu\nu}^1(x)S_{\lambda\rho}^3(y)\right>=
\left<S_{\mu\nu}^2(x)S_{\lambda\rho}^3(y)\right>=
\left<S_{\mu\nu}^1(x)S_{\lambda\rho}^2(y)\right>$$
and the facts that for every $c$, $\left(s_a^{(c)}\right)^2=2$,
$s_1^{(c)}s_2^{(c)}+s_1^{(c)}s_3^{(c)}+s_2^{(c)}s_3^{(c)}=-1$,
we can write

$$s_a^{(c)}s_b^{(c)}\left<S_{\mu\nu}^a(z)S_{\lambda\rho}^b(u)\right>=
2\left(\left<S_{\mu\nu}^1(z)S_{\lambda\rho}^1(u)\right>-
\left<S_{\mu\nu}^1(z)S_{\lambda\rho}^2(u)\right>\right).$$
This leads to the following system of equations:

$$\Delta\hat D\left((x-y)^2\right)=48\pi^2g_m^2\eta^4\int d^4z
\int d^4u D_{m_B}^{(4)}(x-z)D_{m_B}^{(4)}(y-u){\cal G}_2(z-u),$$

$$\Delta\hat G\left((x-y)^2\right)=96\pi^2g_m^2\eta^4\int d^4z
\int d^4u D_{m_B}^{(4)}(x-z)D_{m_B}^{(4)}(y-u)
\left(f_{+}(z-u)-f_{-}(z-u)\right),$$
where $\hat G$ is given by Eq.~(\ref{g}) with the replacement 
${\cal D}_1\to\hat D_1$. Carrying now the integrals out 
analogously to how it was
done in the previous Subsection for the $SU(2)$-case, we get

$$\Delta\hat D\left(x^2\right)=m_B^2\left(D_{M_2}^{(4)}(x)
-D_{m_B}^{(4)}(x)\right),$$

$$\Delta\hat G\left(x^2\right)=4\left[\left(\frac{m_2}{M_2}\right)^2
D_0^{(4)}(x)-D_{m_B}^{(4)}(x)+\left(\frac{m}{M_2}\right)^2
D_{M_2}^{(4)}(x)\right].$$
Together with the old expressions for the functions $\hat D$ and 
$\hat D_1$ (given by Eqs.~(\ref{dvadtri}) and 
(\ref{dvadchetyr}), respectively, with $m\to m_B$),  
which did not account for the screening effect in the 
gas of vortex loops~\footnote{
It is remarkable that these expressions again become exactly cancelled
by the corresponding terms in $\Delta\hat D$ and $\Delta\hat D_1$.}, 
we finally obtain that $\hat D^{\rm full}$
and $\hat D_1^{\rm full}$ are given by Eqs.~(\ref{Dtot}) and 
(\ref{D1tot}), respectively, with the replacements 
$m\to m_B$, $m_D\to m_2$, and $M\to M_2$. 
Therefore the whole discussion, following after Eq.~(\ref{D1tot}),  
remains the same modulo these replacements. In particular, 
when $m_2$ vanishes,
{\it i.e.} one disregards the effect of screening, the old
expressions for the functions $\hat D$ and $\hat D_1$ are recovered.

\section{Modifications of the Propagators of the Dual Vector Bosons 
due to the Screening in the Gas of Vortex Loops}

In the present Section, we shall investigate the influence
of screening of the dual vector bosons in the gas of vortex 
loops to the propagators of these bosons themselves. Let us start with the 
$SU(2)$-case~(\ref{vosem}) by studying the Wilson loop of a test 
magnetic particle, whose charge is in the factor $n$ larger than that
of the dual Higgs field. Such a Wilson loop has the form 

\begin{equation}
\label{magwil}
\left<W(C)\right>=\left<\exp\left(2ig_mn
\int d^4xB_\mu j_\mu\right)\right>\simeq
\exp\left[-2g_m^2n^2\int d^4x\int d^4yj_\mu(x)\left<B_\mu(x)B_\nu(y)
\right>j_\nu(y)\right],
\end{equation}
where $j_\mu(x)=\oint\limits_{C}^{}dx_\mu(s)\delta(x-x(s))$, and 
the average is defined by the partition function~(\ref{vosem}) 
with $F_{\mu\nu}^{\rm e}=0$.
Clearly, in the derivation of the last equality on the R.H.S. of 
Eq.~(\ref{magwil}) we have used the cumulant expansion in the 
bilocal approximation. On the other hand, one can derive the string 
representation for this Wilson loop, and the result reads~\cite{19}:

$$
\left<W(C)\right>=\exp\left[-2g_m^2n^2
\int d^4x\int d^4yj_\mu(x)D_m^{(4)}(x-y)
j_\mu(y)\right]\times$$

$$\times\frac{1}{{\cal Z}}\int {\cal D}x_\mu(\xi)\exp
\left[-(\pi\eta)^2\int d^4x\int d^4y\Sigma_{\mu\nu}(x)
D_m^{(4)}(x-y)\Sigma_{\mu\nu}(y)+i\int d^4xJ_{\mu\nu}\Sigma_{\mu\nu}
\right],$$
where 

\begin{equation}
\label{Jcap}
J_{\mu\nu}(x)\equiv\pi n\varepsilon_{\mu\nu\lambda\rho}\int d^4y
j_\lambda(y)\partial_\rho^x\left[D_m^{(4)}(x-y)-D_0^{(4)}(x-y)\right].
\end{equation}
Using again the bilocal approximation for the average over vortex
loops, we have 

$$
\left<W(C)\right>=\exp\left[-2g_m^2n^2
\int d^4x\int d^4yj_\mu(x)D_m^{(4)}(x-y)
j_\mu(y)-\right.$$

\begin{equation}
\label{wCmo}
\left.-\frac12\int d^4x\int d^4yJ_{\mu\nu}(x)\left<S_{\mu\nu}(x)
S_{\lambda\rho}(y)\right>J_{\lambda\rho}(y)\right],
\end{equation}
where the correlation function of the vortex loops,
$\left<S_{\mu\nu}(x)S_{\lambda\rho}(y)\right>$, is given by 
Eq.~(\ref{SScor}). Inserting now Eqs.~(\ref{SScor}) and (\ref{Jcap})
into Eq.~(\ref{wCmo}), comparing the latter one with Eq.~(\ref{magwil}),
using the conservation of $j_\mu$ and the fact that 
$\partial^2g={\cal G}$, 
we get after straightforward calculations:

$$
\left<B_\mu(x)B_\nu(0)\right>=\delta_{\mu\nu}\Biggl[D_m^{(4)}(x)+
$$

\begin{equation}
\label{BBinterm}
+\frac{g^2}{16}\int d^4z\int d^4u\left(D_m^{(4)}(x-z)-D_0^{(4)}(x-z)
\right){\cal G}(z-u)
\partial^2\left(D_m^{(4)}(u)-D_0^{(4)}(u)\right)
\Biggr].
\end{equation}
Substituting now here the explicit expression~(\ref{calGM}) 
for the function ${\cal G}$ and using the Appendix for the calculation 
of the resulting integrals, we eventually arrive at the following 
simple expression for the propagator of the dual vector boson:

\begin{equation}
\label{BBnew}
\left<B_\mu(x)B_\nu(0)\right>=\delta_{\mu\nu}\frac{1}{M^2}
\left(m_D^2D_0^{(4)}(x)+m^2D_M^{(4)}(x)\right).
\end{equation}
This equation replaces the classical expression 
$\delta_{\mu\nu}D_m^{(4)}(x)$, one has without taking screening 
into account. Clearly, as it should be,
this old expression is reproduced upon taking 
the limit $m_D\to 0$ in Eq.~(\ref{BBnew}).

Let us now consider the $SU(3)$-case~(\ref{suz2}), where the 
magnetic Wilson loop has the form

$$
\left<W(C_1,C_2,C_3)\right>=\left<\exp\left(2ig_mn\int d^4x
{\bf B}_\mu{\bf j}_\mu\right)\right>\simeq
$$

\begin{equation}
\label{su3mag}
\simeq\exp\left[-2g_m^2n^2\int d^4x\int d^4yj_\mu^a(x)
\left<B_\mu^a(x)B_\nu^b(y)\right>
j_\nu^b(y)\right].
\end{equation}
Similarly to Eq.~(\ref{colcur}), we have used here the fact
that the monopole charges are distributed over the root lattice. Therefore,
we have set ${\bf j}_\mu=
{\bf e}_aj_\mu^a$, where $j_\mu^a(x)\equiv\oint\limits_{C_a}^{}
dx_\mu^{(a)}(s)\delta\left(x-x^{(a)}(s)\right)$ with the contour 
$C_a$ parametrized by the vector $x_\mu^{(a)}(s)$. We have also
introduced the notation $B_\mu^a\equiv {\bf B}_\mu{\bf e}_a$ and 
used the cumulant expansion in the bilocal approximation.
In what follows, we shall 
be interested in the expression for the propagator $\left<B_\mu^a(x)
B_\nu^b(0)\right>$. To derive it, let us again consider the string 
representation for the Wilson loop under study. It turns out to 
be analogous to the one we had in the $SU(2)$-case and reads:

$$\left<W(C_1,C_2,C_3)\right>=
\exp\left[-3g_m^2n^2
\int d^4x\int d^4yj_\mu^a(x)D_{m_B}^{(4)}(x-y)
j_\mu^a(y)\right]\times$$

$$\times\frac{1}{{\cal Z}}\int {\cal D}x_\mu^a(\xi)
\delta\left(\sum\limits_{a=1}^{3}\Sigma_{\mu\nu}^a\right)
\exp\left[-(\pi\eta)^2\int d^4x\int d^4y\Sigma_{\mu\nu}^a(x)
D_{m_B}^{(4)}(x-y)\Sigma_{\mu\nu}^a(y)+i\int d^4xJ_{\mu\nu}^a
\Sigma_{\mu\nu}^a\right],$$
where 

\begin{equation}
\label{Jsu3}
J_{\mu\nu}^a(x)\equiv\pi n\varepsilon_{\mu\nu\lambda\rho}\int d^4y
j_\lambda^a(y)\partial_\rho^x\left[D_{m_B}^{(4)}(x-y)-D_0^{(4)}(x-y)\right].
\end{equation}
Applying the cumulant expansion, we have in the bilocal approximation:

$$\left<W(C_1,C_2,C_3)\right>=
\exp\left[-3g_m^2n^2
\int d^4x\int d^4yj_\mu^a(x)D_{m_B}^{(4)}(x-y)
j_\mu^a(y)-\right.$$

\begin{equation}
\label{wccc}
\left.-\frac12\int d^4x\int d^4yJ_{\mu\nu}^a(x)\left<S_{\mu\nu}^a(x)
S_{\lambda\rho}^b(y)\right>J_{\lambda\rho}^b(y)\right],
\end{equation}
where the correlation functions of vortex loops,
$\left<S_{\mu\nu}^a(x)S_{\lambda\rho}^b(y)\right>$, are defined
by Eqs.~(\ref{ssaa}) and (\ref{ssab}). These equations, once being inserted 
into Eq.~(\ref{wccc}) together 
with Eq.~(\ref{Jsu3}), yield upon some calculations 
and comparison of the result with Eq.~(\ref{su3mag}):

$$\left<B_\mu^1(x)B_\nu^1(0)\right>=
\left<B_\mu^2(x)B_\nu^2(0)\right>=$$

\begin{equation}
\label{bbdiag}
=\frac13\delta_{\mu\nu}
\left\{\frac53D_{m_B}^{(4)}(x)+m_2^2\left[
\left(\frac{m_B}{m_1M_1}\right)^2D_{M_1}^{(4)}(x)+
\left(\frac{m_B}{m_2M_2}\right)^2D_{M_2}^{(4)}(x)+
\frac{M_1^2+M_2^2}{M_1^2M_2^2}D_0^{(4)}(x)\right]\right\},
\end{equation}

$$
\left<B_\mu^1(x)B_\nu^2(0)\right>=
$$

\begin{equation}
\label{bboff}
=\frac13\delta_{\mu\nu}\left\{
\frac23D_{m_B}^{(4)}(x)+m_2^2\left[
\left(\frac{m_B}{m_1M_1}\right)^2D_{M_1}^{(4)}(x)-
\left(\frac{m_B}{m_2M_2}\right)^2D_{M_2}^{(4)}(x)-
\frac{2m_2^2}{M_1^2M_2^2}D_0^{(4)}(x)\right]\right\}.
\end{equation}
These equations represent the desired modifications of 
the propagators of the dual vector bosons in the model~(\ref{suz2})
due to the Debye screening of the magnetic charges of these bosons in the 
gas of electric vortex loops. As one can see, in the limit when this 
effect is disregarded, {\it i.e.} $m_1,m_2\ll m_B$, the 
diagonal propagators~(\ref{bbdiag}) go over to the classical expression 
$\delta_{\mu\nu}D_{m_B}^{(4)}(x)$, whereas the off-diagonal one,
(\ref{bboff}), vanishes. This means that the (quantum) effect of screening 
leads in particular to the appearance of the nontrivial correlations between 
dual vector bosons of different types.

\section{Conclusion}

In the present paper, we have investigated
field correlators in the AP $SU(2)$- and $SU(3)$-theories.
These correlators are of the two types. First of them are the 
correlators of electric field strengths, which are the most 
important ones. That is because they correspond to the gauge-invariant
correlators in the real non-Abelian theories, which play the
major r\^ole in the SVM and owing to that are widely used in 
the phenomenological applications. In AP theories, 
such correlators have up to now been evaluated only classically.
In the present paper, we have improved on these calculations
by evaluating the contributions to the correlators, brought about
by the screening of the dual vector bosons in the gas of virtual
vortex loops. This effect is essentially quantum as well as such a 
gas itself. In this way, it has been found that the correlation length
of the vacuum in the models under study becomes modified from the 
inverse mass of the dual vector bosons, those acquire by virtue of 
the Higgs mechanism, to their inverse full mass, which takes also
into account the effect of Debye screening. Besides that, in one of the two
coefficient functions, which parametrize the bilocal correlator of 
electric field strengths within the SVM, there appears also 
a nontrivial power-like IR part, which was absent on the classical 
level. It has also been checked that in the limit when the effect
of screening is disregarded, the obtained novel expressions for the 
bilocal correlators in the AP theories go over to the classical ones,
as it should be. It has also been discussed that the found modifications
of the bilocal correlators do not affect the string tension, 
since this quantity depends only on the {\it v.e.v.} of the dual 
Higgs field and not on the mass of the dual vector bosons. Contrary
to that, the coupling constant of the so-called rigidity term changes
due to the screening.

The correlators of the second kind, we have discussed in this paper,
were the propagators of the dual vector bosons themselves. It turned out
that (contrary to what happened to the correlation length of the vacuum 
in electric correlators) in the magnetic  
propagators the effect of screening does not lead simply to the 
change of the mass from the pure Higgs to the full one. 
There rather appear the expressions quite of a novel form, which,
however, also go over to the classical ones when the effect of 
screening is disregarded. Besides that it has been found that
in the $SU(3)$-case, screening leads also to the appearance 
of the nonvanishing correlations between the fields of the dual
vector bosons of different kinds. This effect is a purely quantum one
and disappears in the classical limit, when the screening is disregarded.

In conclusion, the obtained results shed some light to the 
vacuum structure of the AP theories and give a new field-theoretical 
status to the SVM of QCD.

\section{Acknowledgments}

The author is indebted to Prof. A. Di Giacomo for useful discussions and 
cordial hospitality. He has also benefitted from valuable discussions
with Dr. N. Brambilla and Profs. H.G. Dosch and M.G. Schmidt. 
Besides that, the author is 
greatful to the whole staff of the Quantum Field Theory Division
of the University of Pisa for kind hospitality and INFN for  
financial support.

\section{Appendix. Calculation of the Integral~(\ref{mMint})}

In this Appendix, we shall present some details of calculation 
of the integral~(\ref{mMint}). Firstly, owing to the definition 
of the functions $D_m^{(4)}$ and $D_M^{(4)}$, we have:

$$\int d^4z D_m^{(4)}(z)D_M^{(4)}(z-x)=\int\frac{d^4p}{(2\pi)^4}
\int\frac{d^4q}{(2\pi)^4}\int d^4z\frac{{\rm e}^{ipz}}{p^2+m^2}
\frac{{\rm e}^{iq(z-x)}}{q^2+M^2}=$$

$$=\int\frac{d^4p}{(2\pi)^4}\frac{{\rm e}^{ipx}}{(p^2+m^2)(p^2+M^2)}.$$
Next, this expression can be rewritten as

$$\int\frac{d^4p}{(2\pi)^4}\int\limits_{0}^{+\infty}d\alpha
\int\limits_{0}^{+\infty}d\beta {\rm e}^{ipx-\alpha(p^2+m^2)-
\beta(p^2+M^2)}=\frac{1}{(4\pi)^2}
\int\limits_{0}^{+\infty}d\alpha
\int\limits_{0}^{+\infty}d\beta\frac{{\rm e}^{-\alpha m^2-\beta M^2
-\frac{x^2}{4(\alpha+\beta)}}}{(\alpha+\beta)^2}.\eqno(A.1)$$
It is further convenient to introduce new integration variables
$a\in [0,+\infty)$ and $t\in [0,1]$ according to the formulae
$\alpha=at$ and $\beta=a(1-t)$. Then, the integration over $t$ 
yields for Eq.~(A.1) the following expression:

$$\frac{1}{(4\pi m_D)^2}\int\limits_{0}^{+\infty}
\frac{da}{a^2}{\rm e}^{-\frac{x^2}{4a}}\left({\rm e}^{-am^2}-
{\rm e}^{-aM^2}\right).$$
Such an integral can be carried out by virtue  
of the formula

$$\int\limits_{0}^{+\infty}x^{\nu-1}
{\rm e}^{-\frac{\beta}{x}-\gamma x}dx=
2\left(\frac{\beta}{\gamma}\right)^{\frac{\nu}{2}}K_\nu\left(2
\sqrt{\beta\gamma}\right),~ \Re\beta>0,~ \Re\gamma>0,$$
and the result has the form of Eq.~(\ref{Mmres}) from the main text.

\end{document}